\documentclass{article}
\pdfoutput=1
\usepackage{graphicx}
\usepackage[round]{natbib}
\title{Using CMOS Sensors in a Cellphone for Gamma Detection and Classification}
\author{Joshua J. Cogliati (Joshua.Cogliati@inl.gov), Kurt W. Derr (kurt.derr@inl.gov) \\
 and Jayson Wharton (jayson.wharton@inl.gov)}
\newcommand{\myfigures}[4]{
\begin{figure}[!htbp]
  \centering
  \includegraphics[scale=#4]{#1}
  \caption{ #2}
  \label{#3}
\end{figure}
}

\newcommand{\degree}{${}^\circ$}

\newcommand{\amTFO}{${}^{241}$Am{}}
\newcommand{\seSF}{${}^{75}$Se{}}
\newcommand{\irONT}{${}^{192}$Ir{}}
\newcommand{\csOTS}{${}^{137}$Cs{}}
\newcommand{\coSZ}{${}^{60}$Co{}}

\begin{document}
\maketitle

\section*{Abstract}

The CMOS camera found in many cellphones is sensitive to ionized
electrons.  Gamma rays penetrate into the phone and produce ionized
electrons that are then detected by the camera.  Thermal noise and
other noise needs to be removed on the phone, which requires an
algorithm that has relatively low memory and computational
requirements.  The continuous high-delta algorithm described fits
those requirements.  Only a small fraction of the energy of even the
electron is deposited in the camera sensor, so direct methods of
measuring the energy cannot be used.  The fraction of groups of lit up
pixels that are lines is correlated with the energy of the gamma rays.
This correlation under certain conditions allows limited low
resolution energy resolution to be performed.

\section{Introduction}


Charge-coupled device (CCD) detectors have been used in X-ray
detection and spectroscopy for over a decade.  Complementary
metal-oxide-semiconductor (CMOS) sensors have recently begun to be
used similarly.  Since a cellphone contains a CMOS sensor and
computational ability, the CellRad project \citep{cellrad_faq} was
created to characterize the cellphone's ability as a
gamma radiation sensor.  This project tested both the ability of the
cellphone to operate as a simple dose rate sensor, and---when using
additional processing on the server side---to do low resolution
spectroscopy.  For these tests, the lens is covered with electrical
tape to block the light, but otherwise the only modification to the
cellphone is the addition of software.

\section{Review of Literature}


CCD have been used for X-ray detection and spectroscopy
\citep{rad_detection_knoll,xray_spectroscopy}.  The CCDs are typically
cooled to reduce the generation of dark current
\citep{ccd_tissue_imaging_system}. The amplification of CCD (which is
used to determine the energy in electron volts deposited per pixel)
can be simply measured if access to the raw pixel values are available
\citep{janesick,room_temp_amplification}.  CMOS have not typically
been used because of lower signal-to-noise ratio and lower dynamic
range. However, lower cost and power are two advantages to using CMOS,
and while CMOS technology is improving and there are mitigations for
reducing the noise \citep{noise_characteristics_cmos}.  For X-ray
spectroscopy, for many events, all the energy is deposited in a single
pixel or a group of pixels, and then this energy can be directly
measured \citep{multipixel_ccd_analysis}.

Using the camera in a cellphone to detect radiation has been proposed
\citep{mike_camera}, and several applications are now commercially
available
\citep{gizmag_wikisense,appolicious_gammapix,pervasive_detection}.

\section{Physics Overview}


Radioactive sources produce ionizing radiation every time a single atom
decays into another atom.  This ionizing radiation can come in several
forms including beta (which are high-energy electrons), alpha (which
is a helium nucleus), and gamma (which are high energy photons).  The
beta and alpha particles will be stopped fairly quickly when they
travel through material, but the gamma photons can travel long
distances before being absorbed.

Different isotopes produce different types of ionizing radiation and
different energies of the radiation.  For example \csOTS{} produces
662 keV photons, but \coSZ{} produces 1173 and 1332 keV photons.
Ordinary light is non-ionizing and has much lower energies that are in
the eV range. For example the photons for red light have about 1.77 eV
of energy, and the range goes up to 3.10 eV for violet light photons.

The photons interact in a variety of ways with the material.  The most
important for CellRad are the photoelectric effect and Compton
scattering.  The photoelectric effect is dominate at lower energies.
For example, almost all visible light photons that a material absorbs
will be absorbed by the photoelectric effect.  In the photoelectric
effect, the incoming photon's energy is completely absorbed, and the
energy is transferred to an electron that is ionized (that is,
stripped off of an atom).

Compton scattering is dominate at intermediate energies around 1 MeV.
In Compton scattering, the incoming photon transfers some of its
energy to the electron that it ionizes.  This results in the ionized
electron having only part of the energy of the photon, so the energy
will vary \citep{mayo}.  

The relation between the angle of the electron, the energy of the
electron, and the energy of incoming photon in Compton scattering is:

\begin{equation}
\label{compton_angular}
\frac{  KE_e (E_\gamma  +   m_e c^2)}{E_\gamma \sqrt{KE_e (2 m_e c^2  + KE_e) }} =   cos \phi_e 
\end{equation}

where $KE_e$ is the energy of the electron, $E_\gamma$ is the energy
of the photon, $\phi_e$ is the angle the electron scatters and $2 m_e
c^2$ is approximately 511 keV.  Figure \ref{angle_to_energy_compton}
shows two different distributions of this equation.

\myfigures{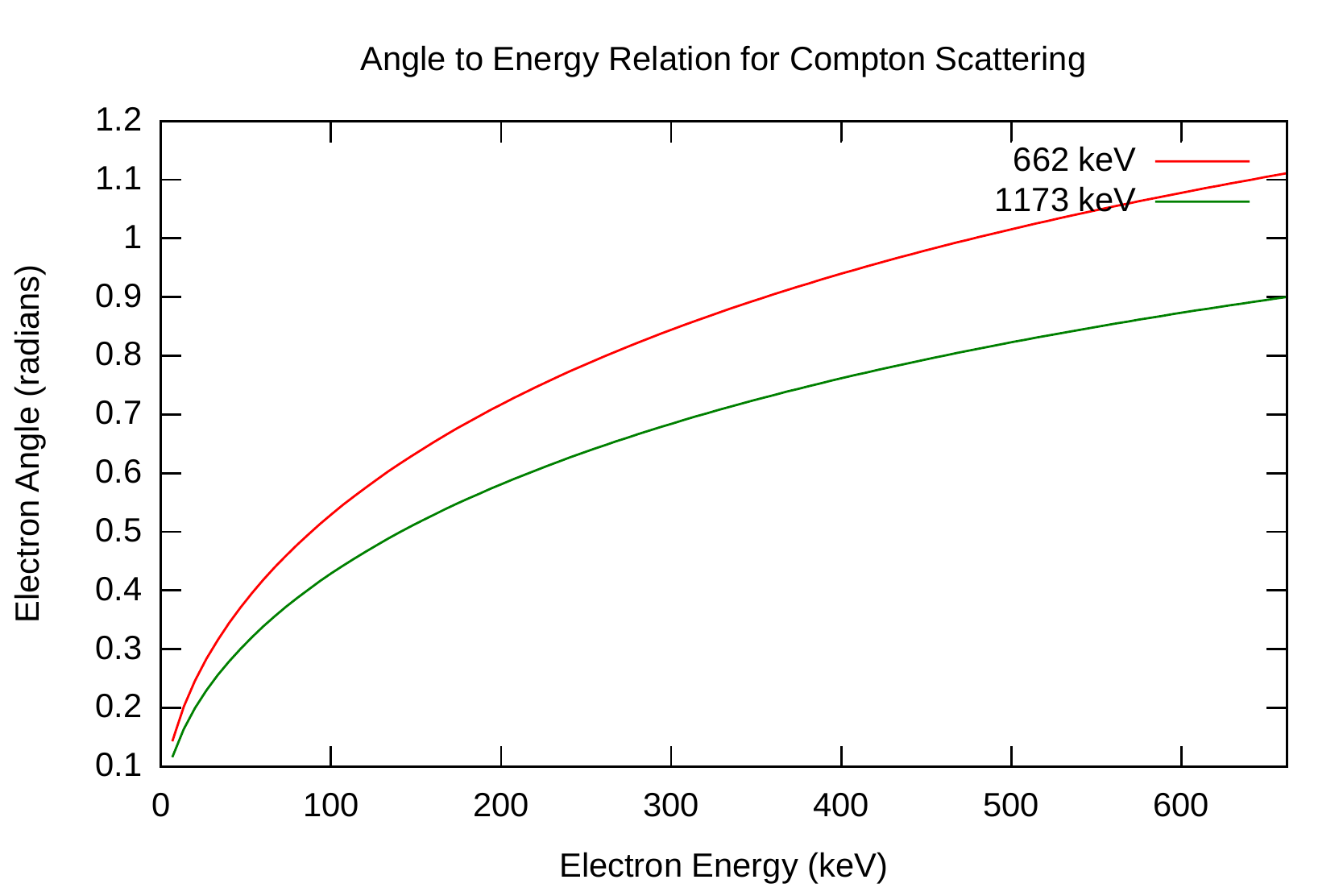}{Compton electron angle-to-energy relations for 662 keV and 1173 keV photons}{angle_to_energy_compton}{0.75}

The ionized electrons may have energy that is much higher than the
energy required to ionize a single electron. In that case, the high
energy electrons will ionize secondary electrons, which will continue
until the energy has been lowered significantly.  A 20 keV electron
might ionize 10,000 secondary electrons.  Electrons with high enough
energies will leave trails of ionization as they travel through the
material.  The higher the energy, the less energy that an electron
will typically leave in a given distance (which is related to stopping
power).

\section{Silicon Surface Barrier Experiments}
\label{ssb_experiments}

\myfigures{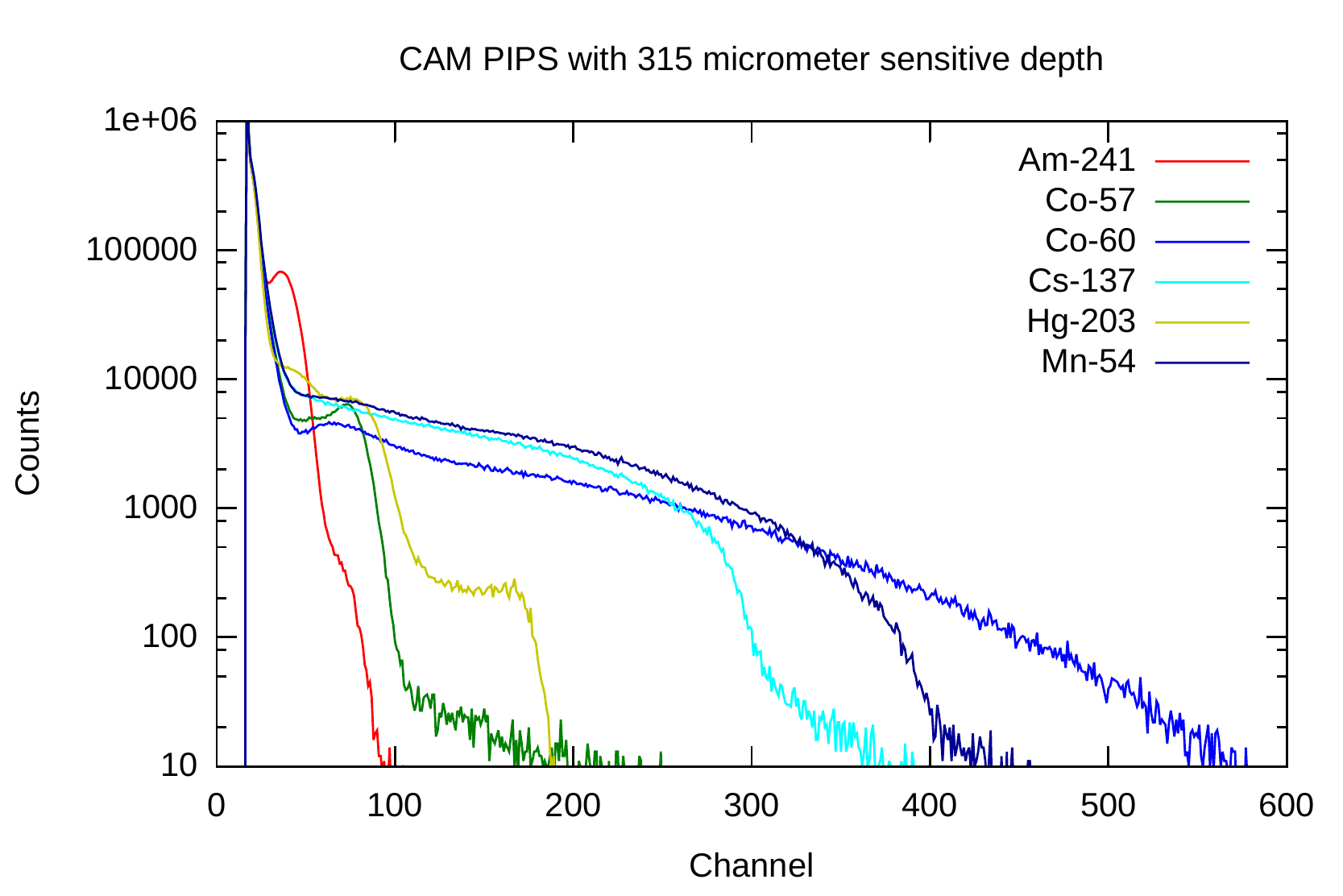}{315 micron ($\mu$m) sensitive depth spectrum}{sb300_spectrum}{0.75}

\myfigures{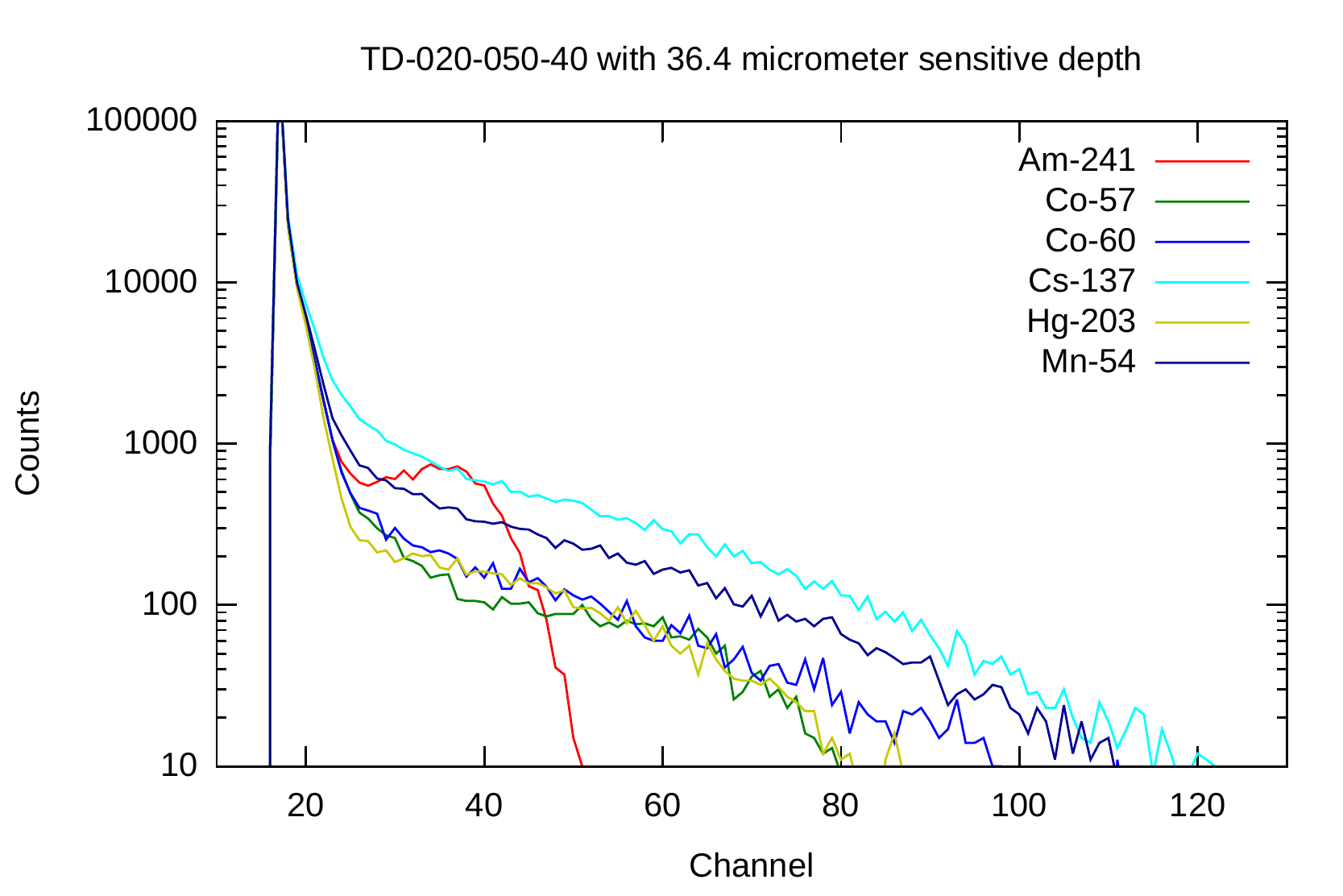}{36.4 micron ($\mu$m) sensitive depth spectrum}{sb45_spectrum}{0.75}

\myfigures{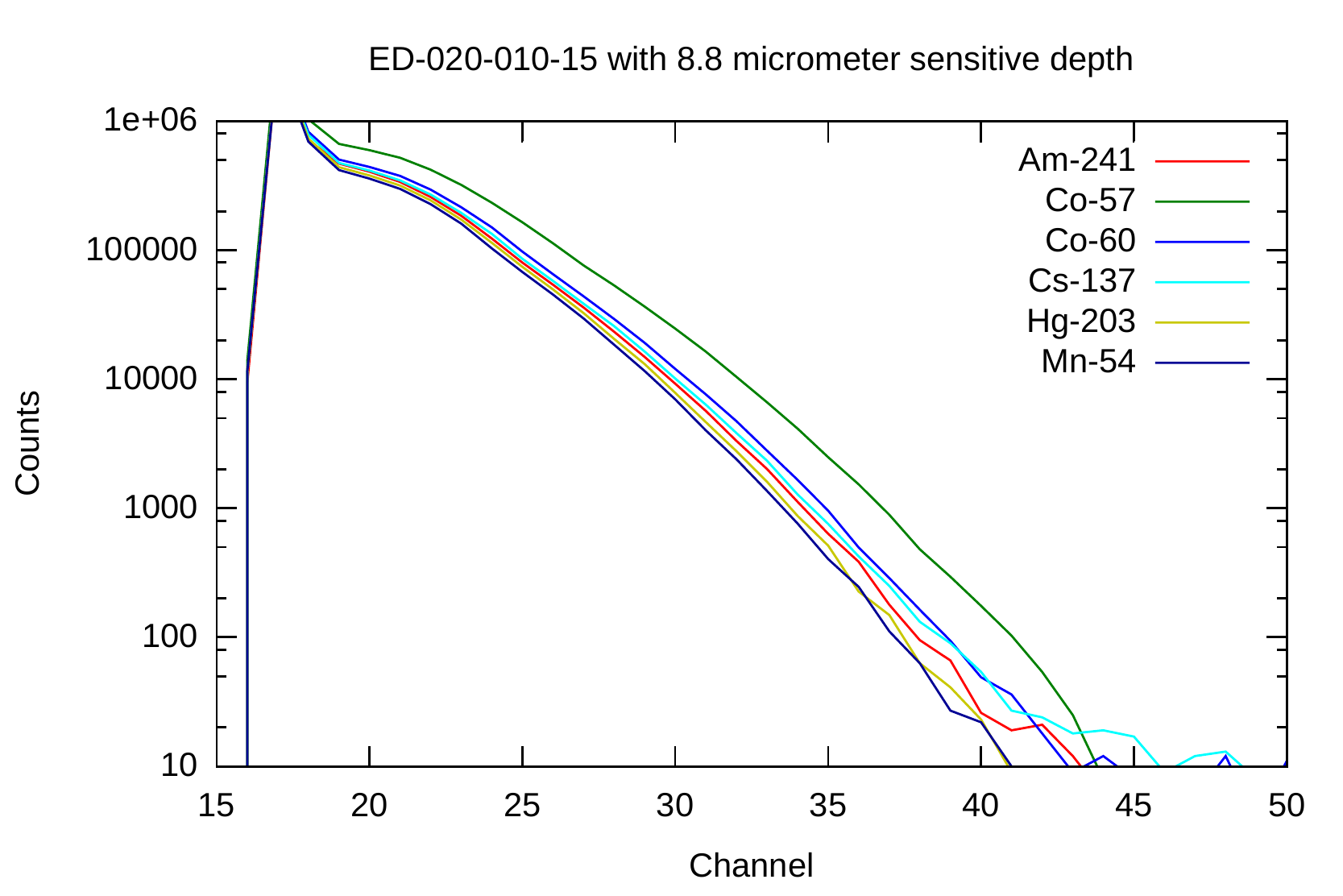}{8.8 micron ($\mu$m) sensitive depth spectrum}{sb15_spectrum}{0.75}

In order to help understand the physics of electrons in silicon,
several experiments were run using silicon surface barrier detectors.
The depth of the sensitive region is not a number that manufacturers
of CMOS cameras provide, but review of the literature gives an
estimate that the thickness is in the range of 10 microns to 0.1
microns \citep{janesick,cmos_imagers}.  Silicon surface barrier
detectors were used as a proxy for a cellphone camera.  Three
detectors were used.  The largest was a CAM Passivated Implanted
Planar Silicon (PIPS) detector produced by Canberra Industries, Inc
which had a sensitive depth (for beta particles) of 315 microns.  The
two smaller surface barrier detectors were produced by Ortec Advanced
Measurement Technology, Inc.  They are both D-series planar totally
depleted silicon surface barrier detectors.  These detectors were
chosen because they are very thin, only 36.4 and 8.8 microns
respectively. The smaller detector is similar in thickness to a pixel
in a digital camera. For all semiconductor radiation detector
measurements, an SMB connector attached the detector signal cable to
an Ortec model 142B pre-amplifier, then on to an Ortec model 572
amplifier and an Ortec model 927 Aspec MCA. The voltage was supplied
by a Canberra model 3102D HV supply.

Even with the thickest detector used in these experiments, energy
spectrum determination is difficult, because of the size of the
detector only part of the energy of the gamma ray remains in the
detector.  Compton edges can be seen in Figure \ref{sb300_spectrum}.
The Compton edge for \amTFO{} at 11 keV can be seen distinctly. However,
as the energy of the gamma ray increases the Compton edges become
smeared out, and the ones at 963 keV and 1117 keV for \coSZ{} are not
clear.  Essentially, the higher the energy of the electron, the higher
the probability that it will exit the detector with significant
energy, instead of depositing the full energy of the Compton scattered
electron.  When the thinner 36.4 micron detector is used, only the
\amTFO{} retains significant structure in Figure \ref{sb45_spectrum}
(Note that the number of channels is decreased in the graph because
less energy was deposited).  With the thinnest detector, there is very
little spectrum information remaining in Figure \ref{sb15_spectrum}
and much smaller amounts of energy deposited.  From this data, direct
measurement of the energy spectrum is expected to be difficult to
impossible with a cellphone camera that has a sensitive region with a
thickness around 10 microns or less.


\section{CMOS Overview}


CCD and CMOS cameras work with visible light by using the
photoelectric effect.  Photons with energies between 1.1 eV to 3.1 eV
ionize one electron creating a single electron-hole pair
\citep{janesick}.  This electron is then captured in the depletion
region of a photodiode or capacitor.  An electric field is used to
separate the electron from the hole in the depletion region, otherwise
they might recombine.  After the collection time has passed, the total
number of electrons is measured and this measurement is used to
determine the brightness for that particular pixel.  For CCD cameras,
the electrons are transferred to a common amplifier and analog to
digital converter.  For CMOS cameras every pixel has at least one
active transistor, so part of the amplification and measurement occurs
at each pixel \citep{cmos_imagers}.  Besides photons causing ionized
electrons, electrons can also be produced by leaky circuits and by
heat.  These electrons result in signal even when there is no light.
These effects tend to be similar from image to image with the same
sensor; when the same sensor is the same temperature, the defects are
even more similar from picture to picture.

For creating colored images, the pixels have different filters put
between the lens and the pixel, which only let a specific color
through.  For example, the first row of pixels might be Red Green R G
R G \ldots and the second row would be Green Blue G B G B \ldots
\citep{ccd_arrays} These individual colors are combined together to
create the full image.  Note that the full image that is returned
usually interpolates the colors to find an individual pixel, so it
includes data from multiple pixels (so the first pixel would get most
of its red data from the physical R, but the G would be interpolated
from the nearby physical pixels).  For detecting radiation, the color
processing makes determining actual energy deposited on the physical
pixels very difficult.  For radiation detection, the camera is covered
so there is no signal from light.  The images also are often
compressed, which for ``lossy'' formats such as JPEG, results in
compression artifacts.  

The gamma rays continually produce ionized electrons.  When there is
an electric field present, the electrons will be collected, but
otherwise they will recombine with holes during the period between
each image collection.

\section{Noise Removal}

Several different preliminary image processing techniques have been
tested to try and remove the non-signal from the images. All these
methods involve looking at multiple images, and processing each
component (red, green, blue) of each pixel individually. These methods
deal with two effects: one is thermal noise, which increases as the
temperature increases, and other is pixels that are defective.  The
defective pixels, or `bad pixels' will show up as bright
points,\footnote{Some defective pixels can show up dark, but these can
  be ignored for radiation detection.} and some of them only light up
periodically, so looking at only two pictures can result in
miscategorizing the bad pixel as actual signal.

The simplest method for noise removal is to take the median value of
the image, and then subtract that from all the images.  This however
will still leave excess noise since some of the noise will be above the
median value.

The next two methods use statistics to try and find the signal.  Some
of the bad pixels do not have constant values, so they can't simply be
subtracted off.  One method is to calculate the standard deviation and
the mean, and then the background image is the highest value that is
less than $\bar{x} + 2 \sigma$:

\begin{equation}
\label{m2sig}
background = \max \left\{ y \in x | y < \bar{x} + 2 \sigma_x \right\}
\end{equation}
where $x$ is the set of pixels at a specific location and specific
component (for example the red component of pixels at 6,3 in all the
images) and $\bar{x}$ in the mean of the set and $\sigma_x$ is the
standard deviation of the set.  A second method is to calculate out a
signal image by using the kurtosis.  The kurtosis is the fourth moment
$\frac{E[(x-E[x])^4]}{E[(x-E[x])^2]^2}-3$, and when the kurtosis is
high, it is quite likely that the image contains ionized signal.  In
this case, the signal is defined as 0 when the kurtosis is below a
threshold, and max - the next highest value otherwise.

The last method to find signal, is called high-delta, and it takes
each pixel in all the images and calculates the max value seen, and
the second highest value seen, and takes the difference between the
two values. This reduces both thermal noise and noise from bad pixels
that periodically lite up.  This processing can be done on either the
phone or the server computer.

\section{On Cellphone Processing}

The cellphone has much more restrictive processing restrictions than
desktop computers.  Every computation done will decrease the battery
life.  The processors on the cellphone are slower than desktop
processors, and there is less memory available, as listed in Table
\ref{cell_phone_data}.  However, transmitting pictures to a central
server for processing requires excessive bandwidth, so preliminary
processing should be done on the phone.  The software is written for
Android 4.0 and higher phones.  The scene mode and the exposure
compensation were chosen to maximize the exposure time and are listed
in Table \ref{dose_to_number_of_groups_phone_list}. The exposure time
was read from the EXIF data.

\begin{table}[!htb]
  \centering
  \caption{Cell phone data}
  \label{cell_phone_data}
  \begin{tabular}{l|l|l|l}
    Phone  & Exposure & Memory & Processor\\
         & Time     &        &         \\
    \hline
    Nexus S  & 1.0 second & 512 MB & 1GHz Cortex A8\\
    Nexus Galaxy &  0.120 second & 1 GB & 1.2 GHz dual-core ARM Cortex-A9 \\
    Nexus 4 & 1.0 second & 2 GB &  	1.5 GHz quad-core Krait\\
  \end{tabular}
\end{table}

The general task of the phone is to filter out various noise sources,
and determine an approximate dose.  The filtering is done by a
continuously running version of the high-delta method. The filtering
is looking at each component of each pixel, and keeping track of the
highest value and the next highest value, and using the difference as
the signal. The dose rate is calculated by using a multiplier of the
number of pixels that are over a signal threshold value.  

The algorithm starts off by creating two arrays of the data (actually
stored in bitmaps since three-dimensional arrays in Java are memory
intensive) for storing the maximum value seen (max) and the second
highest value (max\_1). These arrays are needed to determine the delta
between the two arrays at each pixel component. Pseudocode for initializing max
and max\_1:

\begin{verbatim}
byte[width,height,pixel_components] max[:,:,:] = 0
byte[width,height,pixel_components] max_1[:,:,:] = 0
\end{verbatim}

For each additional picture the arrays are updated.  The algorithm
checks to see if the current value is greater than what has been seen
at that pixel and component, and if so calculates the signal and
updates the arrays.  The dose is a simple multiplication of the number
of pixel components found over a signal threshold. The pseudocode for
updating max and max\_1 data from image and calculating
overSignalThreshold and the signal are:

\begin{verbatim}
overSignalThreshold = 0
maxSignal = 0
for x in image_width:
   for y in image_height:
      for c in pixel_components:
         value = image[x,y,c]
         max_1 = max_1_image[x,y,c]
         max = max_image[x,y,c]
         if value > max_1:
            max_1 = value
            if max_1 > max: 
               swap(max_1,max)
            signal = value - max_1
            maxSignal = max(signal,maxSignal)            
            if signal > signalThreshold:
               overSignalThreshold += 1
            max_1_image[x,y,c] = max_1
            max_image[x,y,c] = max
pictureDose = ovstToDoseMult*overSignalThreshold
\end{verbatim}

Every so often (twenty images or forty images are commonly used in
CellRad), a combined image is generated.  The combined image has much
less noise than each individual image.  The combinedDose will be more
accurate than the pictureDose calculated on each individual picture
for two reasons. First, it combines data from multiple pictures, which
will decrease statistical variance of the actual dose because it is
mostly\footnote{If some of the pixels of signal overlap, then this
  will not truly be the sum of the individual pictures.} the sum of
independent counts. Secondly, the high delta formula is operating over
the full number of combined pictures instead of only having partial
data when the algorithm has zeroed max\_1.  The max array is
reinitialized from the max\_1 array.  The max array needs to be reset
because it contains the signal data, which if it was never reset would
gradually fill with the highest pixel component value seen.
Conversely, if both max and max\_1 were zeroed then the information
about current noise levels would need to be rediscovered (which would
mess up the current pictureDose until enough pictures had been taken).
Current noise levels vary because the temperature of the CMOS sensor
varies. Pseudocode for creating a picture (combined\_image) from max
and max\_1 data and updating:

\begin{verbatim}
overSignalThreshold = 0
maxSignal = 0
for x in image_width:
   for y in image_height:
      for c in pixel_components:
         signal = max[x,y,c] - max_1[x,y,c]
         combined_image[x,y,c] = signal
         maxSignal = max(signal,maxSignal)            
         if signal > signalThreshold:
            overSignalThreshold += 1
combinedDose = ovstToDoseMult*overSignalThreshold/numCombinedPictures

max[:,:,:] = max_1[:,:,:]
max_1[:,:,:] = 0
\end{verbatim}

At the Idaho National Laboratory's (INL) Health Physics Instrument
Laboratory (HPIL), three different cellphone model's were placed in
gamma fields produced by \coSZ{} or \csOTS{} sources.  The cellphone's
back camera was facing the source and the lens was covered with
electrical tape.  Forty pictures were recorded for each different dose
and nuclide test.  The data was then processed using the on cellphone
software.  Threshold values and
over-threshold-to-dose-mult values were chosen to minimize false
signal and give correct dose results with a 100 mrem/hr \csOTS{} field
and are given in Table \ref{cell_phone_filter_settings}.  Results from
individual phones are given in Tables
\ref{nexus_s_on_phone_calculated},
\ref{nexus_galaxy_on_phone_calculated}, and
\ref{nexus_4_on_phone_calculated}.  From these tables the accuracy of
the dose calculated can be compared to the actual dose.

\begin{table}[!htb]
  \centering
  \caption{Cell phone filter settings}
  \label{cell_phone_filter_settings}
  \begin{tabular}{l|l|l|l}
    Phone & Signal  & Threshold & OVST To\\
        & Threshold &           & Dose Mult \\
    \hline
    Nexus S & 60 & 110 & 0.141\\
    Nexus Galaxy & 60 & 110 &  2.33\\
    Nexus 4 & 32 & 32 & 1.129\\
  \end{tabular}
\end{table}

\begin{table}[!htb]
  \centering
  \caption{Nexus S On Phone Calculated Data in mrem/hr}
  \label{nexus_s_on_phone_calculated}
  \begin{tabular}{r|l|r|r|r|r|r}
    Actual Dose	&nuclide	&Pictures	& Average Dose& StdDev	& Min	&Max\\
    \hline
    0	&NA	&39	&0.00	&0.00	&0.00	&0.00	\\
    1	&\csOTS{}	&41	&0.59	&1.89	&0.00	&9.58	\\
    10	&\csOTS{}	&40	&10.91	&15.26	&0.00	&71.50	\\
    100	&\coSZ{}	&40	&65.65	&35.40	&18.88	&165.45	\\
    100	&\csOTS{}	&40	&100.74	&34.98	&41.04	&178.18	\\
    1000	&\coSZ{}	&40	&1048.65	&125.70	&816.53	&1324.04\\
    1000	&\csOTS{}	&41	&1043.84	&120.14	&734.02	&1316.74\\
    10000	&\coSZ{}	&40	&9482.18	&670.36	&8291.71	&10729.29\\
    10000	&\csOTS{}	&40	&9944.66	&807.68	&8333.61	&11552.83\\
    100000      &\coSZ{}  &40     &58918.20       &25080.20        &0.00   &126870.18\\
    100000	&\csOTS{}	&40	&54614.10	&16064.60	&30871.41	&94148.34\\
  \end{tabular}
\end{table}

For the Nexus S, the average dose rate is with 10\% for most of the
higher dose rates.  For the 100,000 mrem/hr dose rate, the high-delta
algorithm began saturating, because approximately 4\% of the pixels in
the image were signal, so in a sequence of images, the max and even
the max\_1 arrays begins to fill with actual gamma signal.  This
decreases the calculated dose since the signal can only be picked up
in the regions where there is no overlap between previous signal and
new signal. Robustly detecting this saturation issue has not been
solved.  The Nexus Galaxy and Nexus 4 do not have this issue in the
dose rate ranges simply because they are less sensitive.  Conversely,
they have problems detecting any dose for the lower dose ranges, and
many of the pictures find zero signal.

\begin{table}[!htb]
  \centering
  \caption{Nexus Galaxy On Phone Calculated Data in mrem/hr}
  \label{nexus_galaxy_on_phone_calculated}
  \begin{tabular}{r|l|r|r|r|r|r}
    Actual Dose	&nuclide	&Pictures	& Average Dose& StdDev	& Min	&Max\\
    \hline
    0	&NA	&40	&0.29	&1.84	&0.00	&11.65\\
    1	&\csOTS{}	&40	&0.00	&0.00	&0.00	&0.00\\
    10	&\csOTS{}	&40	&12.64	&38.69	&0.00	&186.40\\
    100	&\coSZ{}	&40	&66.46	&127.50	&0.00	&507.94\\
    100	&\csOTS{}	&40	&98.03	&160.38	&0.00	&582.50\\
    1000	&\coSZ{}	&40	&641.45	&396.17	&100.19	&1908.27\\
    1000	&\csOTS{}	&40	&771.00	&398.74	&0.00	&1812.74\\
    10000	&\coSZ{}	&40	&6616.27	&1405.21	&3341.22	&9648.53\\
    10000	&\csOTS{}	&40	&7276.30	&1394.52	&4678.64	&10478.01\\
    100000	&\coSZ{}	&40	&63945.60	&6016.57	&54899.46	&82696.36\\
    100000	&\csOTS{}	&40	&68712.50	&4693.94	&60326.03	&79508.92\\
  \end{tabular}
\end{table}

\begin{table}[!htb]
  \centering
  \caption{Nexus 4 On Phone Calculated Data in mrem/hr}
  \label{nexus_4_on_phone_calculated}
  \begin{tabular}{r|l|r|r|r|r|r}
    Actual Dose	&nuclide	&Pictures	& Average Dose& StdDev	& Min	&Max\\
    \hline
    0	&NA	&39	&0.00	&0.00	&0.00	&0.00	\\
    1	&\csOTS{}	&40	&1.98	&12.50	&0.00	&79.03	\\
    10	&\csOTS{}	&40	&9.94	&32.80	&0.00	&162.58	\\
    100	&\coSZ{}	&39	&77.61	&100.99	&0.00	&410.96	\\
    100	&\csOTS{}	&37	&105.33	&93.72	&0.00	&343.22	\\
    1000	&\coSZ{}	&39	&807.21	&315.44	&189.67	&1756.72\\
    1000	&\csOTS{}	&39	&1128.74	&339.84	&461.76	&1801.88\\
    10000	&\coSZ{}	&39	&8219.99	&869.95	&6833.84	&10778.56\\
    10000	&\csOTS{}	&39	&10609.10	&1139.78	&8512.66	&14048.15\\
    100000	&\coSZ{}	&39	&83476.70	&6870.55	&72193.90	&97361.57\\
    100000	&\csOTS{}	&39	&102907.00	&8178.31	&84725.80	&118119.36\\
  \end{tabular}
\end{table}

Table \ref{battery_drain_time} lists the time it takes to drain the
battery from full to 9\%.  At 9\% battery picture taking stops.
Pictures are taken either every 60 seconds, every 20 seconds or every
10 seconds.  For both tests, combined images were created and sent
every 40 pictures.  The filter time is the average amount of time that
the software spends running the noise filtering algorithms.

\begin{table}[!htb]
  \centering
  \caption{Time to drain battery from full to 9\%}
  \label{battery_drain_time}
  \begin{tabular}{l|l|l|l|l}
    Phone & 60 second interval & 20 second interval & 10 second interval & filter time\\
    \hline
    Nexus S & 439 minutes & 300 minutes & 209 minutes & 1478 ms\\
    Nexus Galaxy & 1070 minutes & 471 minutes & 399 minutes & 1098 ms\\ 
    Nexus 4 & 2520 minutes & 609 minutes & 556 minutes & 1250 ms\\ 
  \end{tabular}
\end{table}

\section{Low Resolution Spectrum Processing}

Traditional spectroscopy cannot be used by the cellphone camera.  The
X-rays are severely attenuated or blocked by the lens. The gamma rays
that do reach the camera sensor deposit only a portion of their energy
in the sensor since the stopping range greatly exceeds the thickness
of the depletion region and the physical pixel size as seen in Section
\ref{ssb_experiments}.  For the Nexus S back camera, the planar pixel
size is approximately 2 microns ($\mu$m) square, based on the outside
dimensions of the camera unit (less than 8 mm x 8 mm).  The small
detector size precludes traditional spectroscopy, but the two
dimensional planar array of pixels allows much different processing to
attempt to extract some of the spectrum data.


On the server, once the image has been processed to remove noise, the
software processes the image to find lines and groups.  The
line\_finder program uses ImageMagick to read in the image files.  The
program uses the average of the red and blue pixels to be the value
used for processing.  The line\_finder program starts with an image
such as Figure \ref{combined_image}.

\myfigures{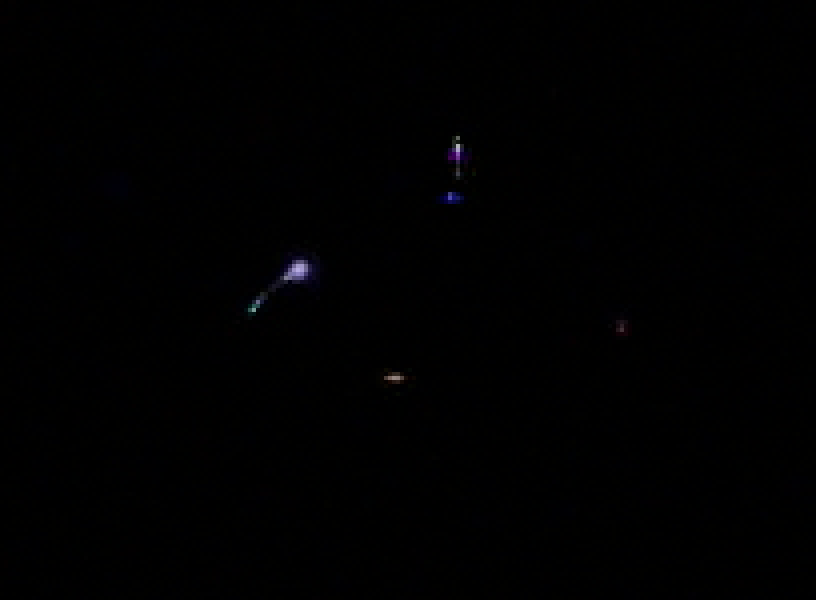}{Portion of combined image from Nexus S in \csOTS{} 100 mrem/hr field}{combined_image}{0.3,natwidth=816,natheight=600}

\myfigures{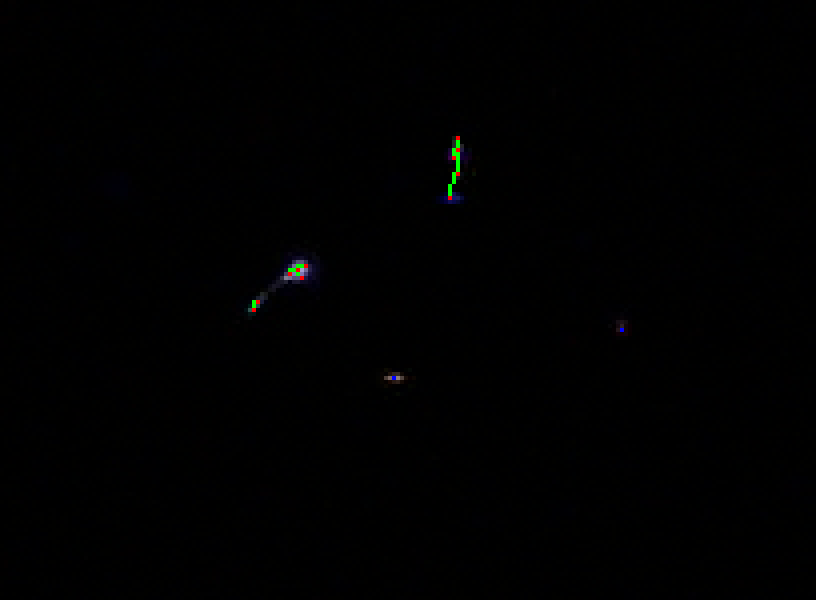}{Line finder processing of portion of combined image from Nexus S in \csOTS{} 100 mrem/hr field}{combined_image_debug}{0.3,natwidth=816,natheight=600}

Two values are used for determining if something is an event or not
and its type.  These are the bright threshold and the bad\_bright
value.  In order for the program to decided that some set of pixels
are an event, at least one pixel needs to be over the bright
threshold.  In order for two bright pixels to be considered part of
the same event, the line between them needs to have an average value
that is higher than the bad\_bright value.  

The value of the bright and bad\_bright can either be specified
through command line arguments, or they can be automatically
determined.  The automatically determined bright value is $\bar{x} +
5\sigma+16$ and the automatically determined $\bar{x}+3\sigma+16$
where $\bar{x}$ is the average pixel value and $\sigma$ is the
standard deviation.


After determining the bright threshold, the program generates a list
of bright points in the image. The points are chosen so that they are
at least the exclude size of 2 pixels away from each other.  The
points are also adjusted within a few pixels to make sure that the
brightest local pixel is chosen.


Next, the program calculates spanning trees between the bright pixels.
For each iteration of the spanning tree algorithm, the program starts
with a bright pixel that is currently unattached to a spanning tree.
The remaining unattached bright pixels are iterated over, and the
closest is attached to the spanning tree if it is closer than the
maximum span length.  If the closest to the currently attached set is
greater than the maximum span length then the current spanning tree is
done, and the next spanning tree will be worked on.  The typical
maximum span length is 10 pixels.

The program calculates statistics for each spanning tree, which are
called groups.  These statistics include the minimum and maximum x and
y coordinates, the average brightness, the brightest pixel and the sum
of the bright pixels.  On a raw image, the sum of the bright pixels is
related to the energy deposited, but on a color image these are only
approximately related.  

For each spanning tree, the program enumerates all the paths from one
end to another end.  For example, if the spanning tree ended up like a
Y, there would be three ends and three paths.  Each of these paths is
a line.  Statistics are printed for each line, such as total length
and start and end locations.

For example, the line\_finder program found five groups, three of
which had lines in Figure \ref{combined_image}.  There were seven
total lines in the image, and the debug output image is Figure
\ref{combined_image_debug}, which shows the line segments in green,
bright pixels that are part of a line in red, and bright pixels that
are unattached in blue.  The ratio of total lines to groups is
correlated with the energy of the photon.  The fraction of groups that
have lines is also correlated to average incoming photon energy.


At the end, combined statistics are outputted such as total number of
lines and total number of groups found and the number of groups that
have lines.

\section{Low Resolution Spectrum Results}


\begin{table}[!htb]
  \centering
  \caption{Most common gamma energies produced by tested nuclides \citep{los_alamos_rad_monitoring}}
  \label{gamma_energies}
  \begin{tabular}{r|p{8cm}}
    Nuclide & Energies (keV) \\
    \hline
    \amTFO{} & 60 (35.9\%), 26 (2.4\%), 33 (0.1\%) \\
    \seSF{} & 265 (59.8\%), 136 (59.2\%), 11 (47.5\%) 280 (25.2\%), 12 (7.3\%), 1
(0.9 \%)\\
    \irONT{} &  317~(82.85\%), 468~(48.1\%), 308~(29.68\%), 296~(29.02\%),67~(4.52\%),  9~(4.1\%), 65~(2.63\%), 76~(1.97\%) \\
    \csOTS{} & 662~(89.98\%), 32~(5.89\%), 36~(1.39\%), 5~(1\%)\\
    \coSZ{} & 1173~(100\%), 1332~(100
  \end{tabular}
\end{table}


Several different curie level sources were tried at the INL site in
August of 2012.  The sources tried were a 1.3 Ci \amTFO{} source, a
1.3 Ci \csOTS{} source (1.6 Ci at 7/8/2003), a 0.1 Ci \coSZ{} source
(0.3 Ci at 7/8/2003), a 0.7 Ci \seSF{} source (62.0 Ci at 6/21/2010)
and a 1.3 Ci \irONT{} source (144.7 Ci at 2/23/2011).  The common
gamma energies are listed in Table \ref{gamma_energies}. These were
tried with the back camera facing the source, but with different
distances ranging from 30 cm to 90 cm to the sources.  For some of the
experiments with the \amTFO{} source, the \csOTS{} source and the
\coSZ{} sources, steel plates up to 0.25 inches thick (0.635 cm) were
put between the source and the cellphone to check the effect of
shielding.  For each single experiment count, 100 pictures were taken
and then combined with the high-delta method, before being analyzed to
determine groups and groups with lines fractions.  Figures
\ref{2012_aug_16_site_cs-137_long_1_bk_hd_64c},
\ref{2012_aug_16_site_co-60_long_1_bk_hd_64c},
\ref{2012_aug_16_site_am-241_long_1_hd_64c},
\ref{2012_aug_16_site_ir-192_long_1_hd_64c}, and
\ref{2012_aug_16_site_se-75_long_1_hd_64c} show example images from
different sources.

The results from 138 experiments are in Figure
\ref{gwl_over_g_site_2012_aug} and the statistics from them in Table
\ref{fraction_of_groups_with_lines_stats}.  From the data using the
fraction of groups with lines, \coSZ{} and \csOTS{} can be clearly
distinguished from each other and from \seSF{} and \amTFO{}.  \irONT{}
overlaps with \csOTS{}, and \seSF{} and \amTFO{} overlap.  The
fraction of groups with lines decreases as the average energy of the
gamma rays increases.  Figure \ref{gwl_to_energy} shows the
relationship between the average energy of the nuclide and the
fraction of groups with lines.

\begin{table}[!htb]
  \centering
  \caption{Fraction of Groups with Lines Statistics}
  \label{fraction_of_groups_with_lines_stats}
  \begin{tabular}{l|l|l|l|l|l}
    &\seSF{}  &\irONT{} &\csOTS{} &\coSZ{}  &\amTFO{}\\
    \hline
    Mean    &0.659  &0.578  &0.527  &0.415  &0.717\\
    Standard Deviation      &0.009  &0.013  &0.028  &0.021  &0.043\\
    Minimum &0.642  &0.557  &0.474  &0.370  &0.625\\
    Maximum &0.668  &0.594  &0.564  &0.447  &0.833\\
    Count   &6      &6      &39     &51     &36\\
  \end{tabular}
\end{table}

\myfigures{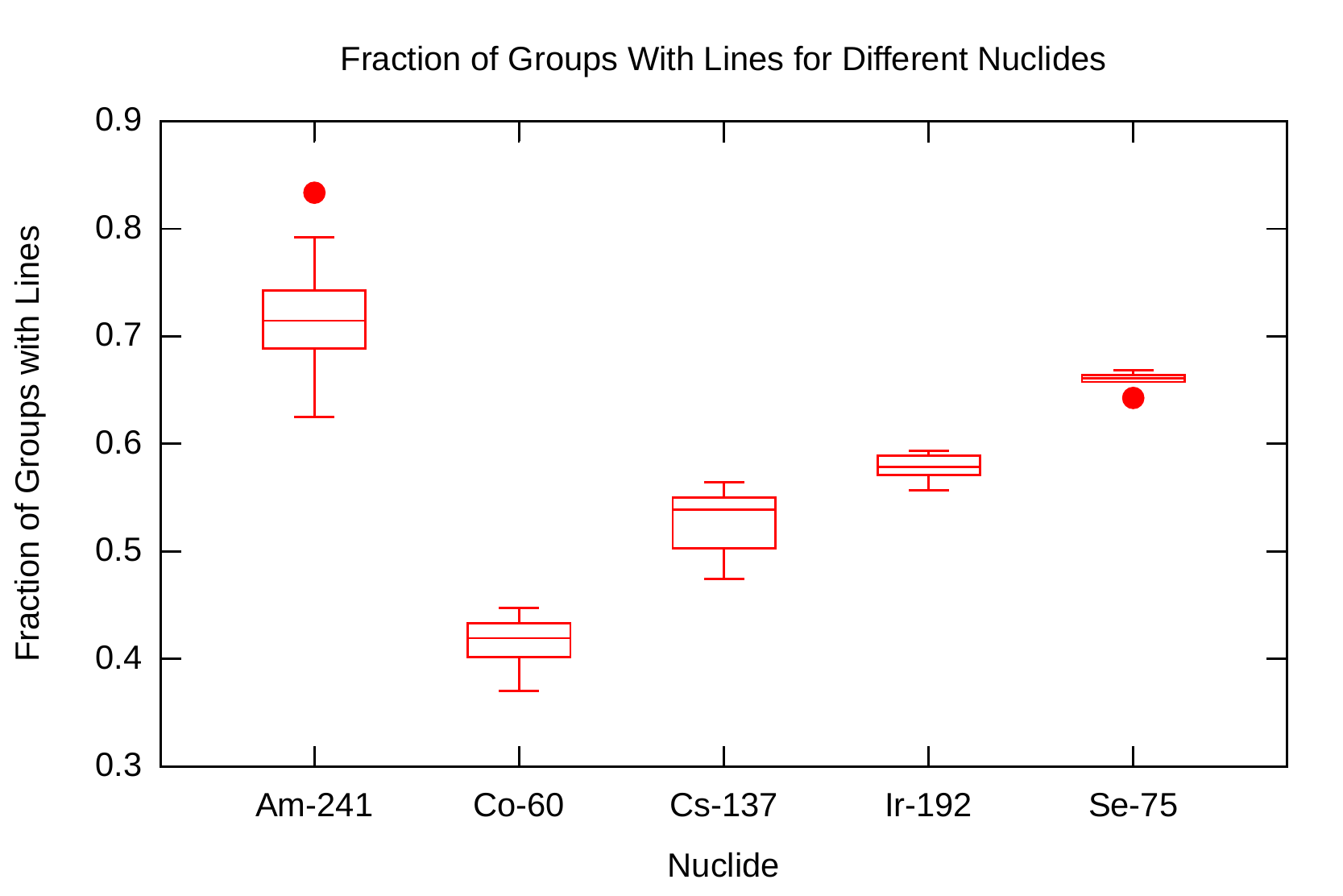}{Nexus S, 1 foot to 3 feet away from various curie level sources}{gwl_over_g_site_2012_aug}{0.75}

\myfigures{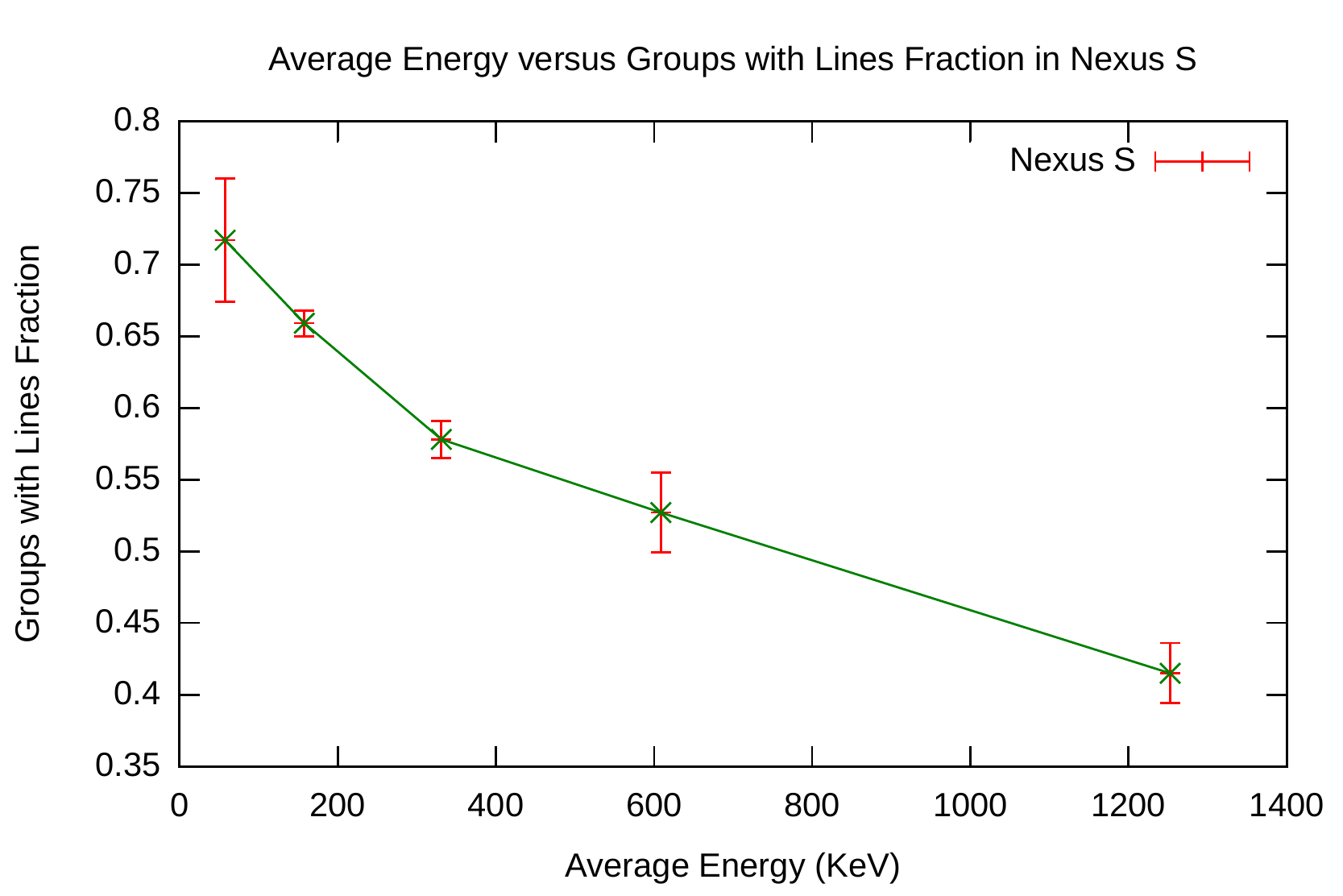}{Comparison of average nuclide energy with fraction of groups with lines and standard deviation error bars}{gwl_to_energy}{0.75}

\section{Rotation Results}


Changing the angle of the incoming gamma rays changes what materials
the gamma rays travel through to get to the sensor.  For example, at
certain angles the gamma rays will travel through the battery before
reaching the camera. This material can attenuate the gamma rays.  The
interaction also can generate high energy electrons that are detected
by the CMOS sensor. From Equation \ref{compton_angular} the electrons
scattered from Compton scattering have an angular energy
dependence. These two effects changes both the number of events that
are seen and the types of features that are seen.  A Nexus S phone was
placed in a 10 R/hr field generated from \csOTS{} and a separate 10
R/hr field generated from \coSZ{} at the INL's HPIL.  The phone was
placed vertically, either taped to a cardboard box with the top of the
phone and the camera above the box as in Figure \ref{nexus_s_plain} or
with the back camera against a 30.5 x 30.5 x 15.2 cm (12 x 12 x 6
inch) block of Lucite (Polymethyl methacrylate) as in Figure
\ref{nexus_s_against_lucite}.

\begin{table}[!htb]
  \centering
  \caption{Nexus S Rotation Data Back Camera}
  \label{nexus_s_rotation_data} 
  \begin{tabular}{r|r|r|r|r|r|r|r|r|r}
    \multicolumn{2}{c}{ }	&\multicolumn{3}{c}{Back Camera Lucite}	&	&\multicolumn{3}{c}{Back Camera Plain}	&\\
    Nuclide	&Degrees&Groups	&GwL	&GwLL	&MLL	&Groups	&GwL	&GwLL	&MLL\\
    \hline
    \csOTS{}	&0	&33699	&15422	&3415	&69.69	&78143	&32324	&7676	&79.40\\
    \csOTS{}	&90	&57803	&28188	&7041	&99.27	&52068	&25246	&6335	&86.24\\
    \csOTS{}	&180	&78089	&33335	&7789	&75.51	&72336	&29394	&7070	&99.62\\
    \csOTS{}	&270	&58905	&28665	&7044	&97.13	&58223	&28111	&7081	&80.72\\
    \hline
    \coSZ{}	&0	&52591	&18649	&4004	&70.88	&99868	&32826	&7383	&115.07\\
    \coSZ{}	&90	&75730	&30407	&7298	&83.80	&73267	&29285	&7064	&128.98\\
    \coSZ{}	&180	&91483	&31471	&7155	&89.08	&88264	&29075	&6494	&102.07\\
    \coSZ{}	&270	&79504	&31720	&7598	&100.65	&75897	&29807	&7117	&106.89\\
  \end{tabular}
\end{table}

\begin{table}[!htb]
  \centering
  \caption{Nexus S Rotation Data Front Camera}
  \label{nexus_s_rotation_data_front} 
  \begin{tabular}{r|r|r|r|r|r|r|r|r|r}

    \multicolumn{2}{c}{ }	&\multicolumn{3}{c}{Front Camera Lucite}	&	&\multicolumn{3}{c}{Front Camera Plain}	&\\
    Nuclide	&Degrees&Groups	&GwL	&GwLL	&MLL	&Groups	&GwL	&GwLL	&MLL\\
    \hline
    \csOTS{}	&0	&254	&92	&6	&21.67	&515	&165	&7	&22.61\\
    \csOTS{}	&90	&445	&150	&16	&20.54	&433	&126	&9	&19.30\\
    \csOTS{}	&180	&609	&184	&17	&17.06	&534	&165	&18	&18.06\\
    \csOTS{}	&270	&435	&127	&12	&17.46	&356	&111	&8	&25.21\\
    \hline
    \coSZ{}	&0	&245	&78	&3	&15.39	&458	&118	&13	&25.32\\
    \coSZ{}	&90	&437	&117	&8	&19.34	&395	&95	&4	&15.86\\
    \coSZ{}	&180	&497	&130	&10	&21.11	&454	&109	&8	&16.39\\
    \coSZ{}	&270	&447	&133	&10	&14.57	&368	&98	&6	&14.53\\
  \end{tabular}
\end{table}

\myfigures{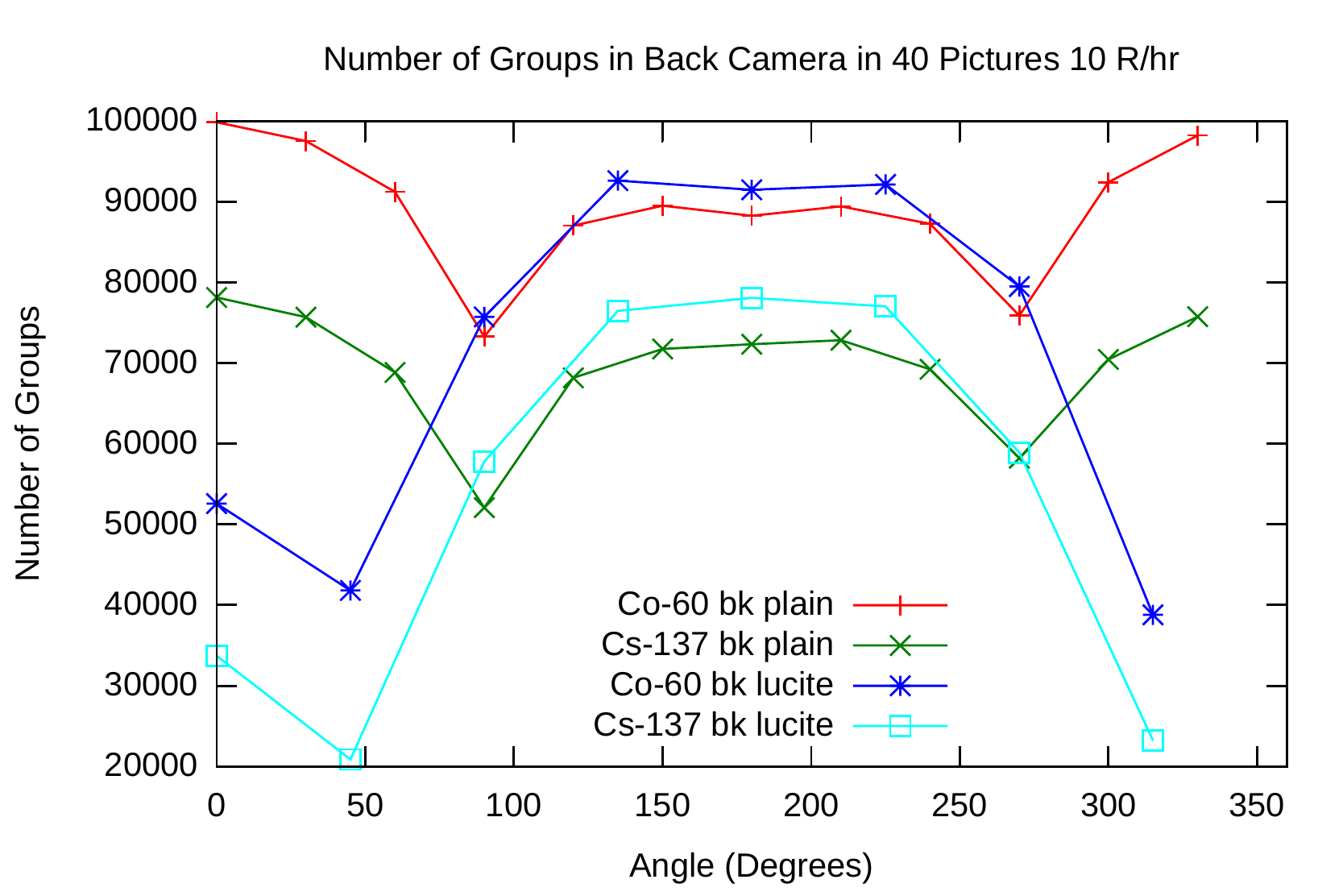}{Nexus S Angular Groups 10 R/hr, 40 pictures back camera 1.0 sec exposure}{number_of_groups_nexus_s_rotation_back_2013_jan_hpil}{0.74}

\myfigures{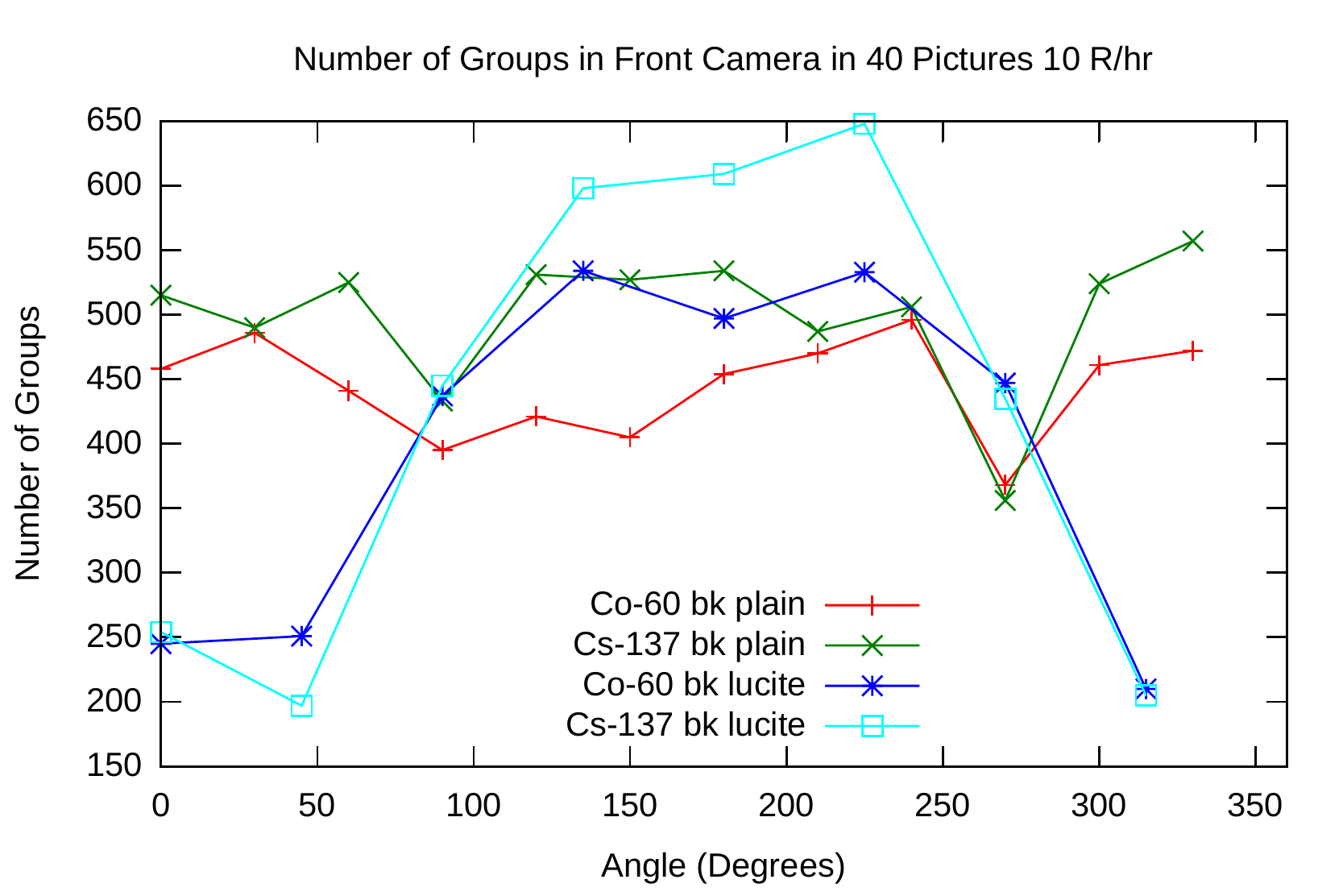}{Nexus S Angular Groups 10 R/hr, 40 pictures front camera 0.125 sec exposure}{number_of_groups_nexus_s_rotation_front_2013_jan_hpil}{0.74}

\myfigures{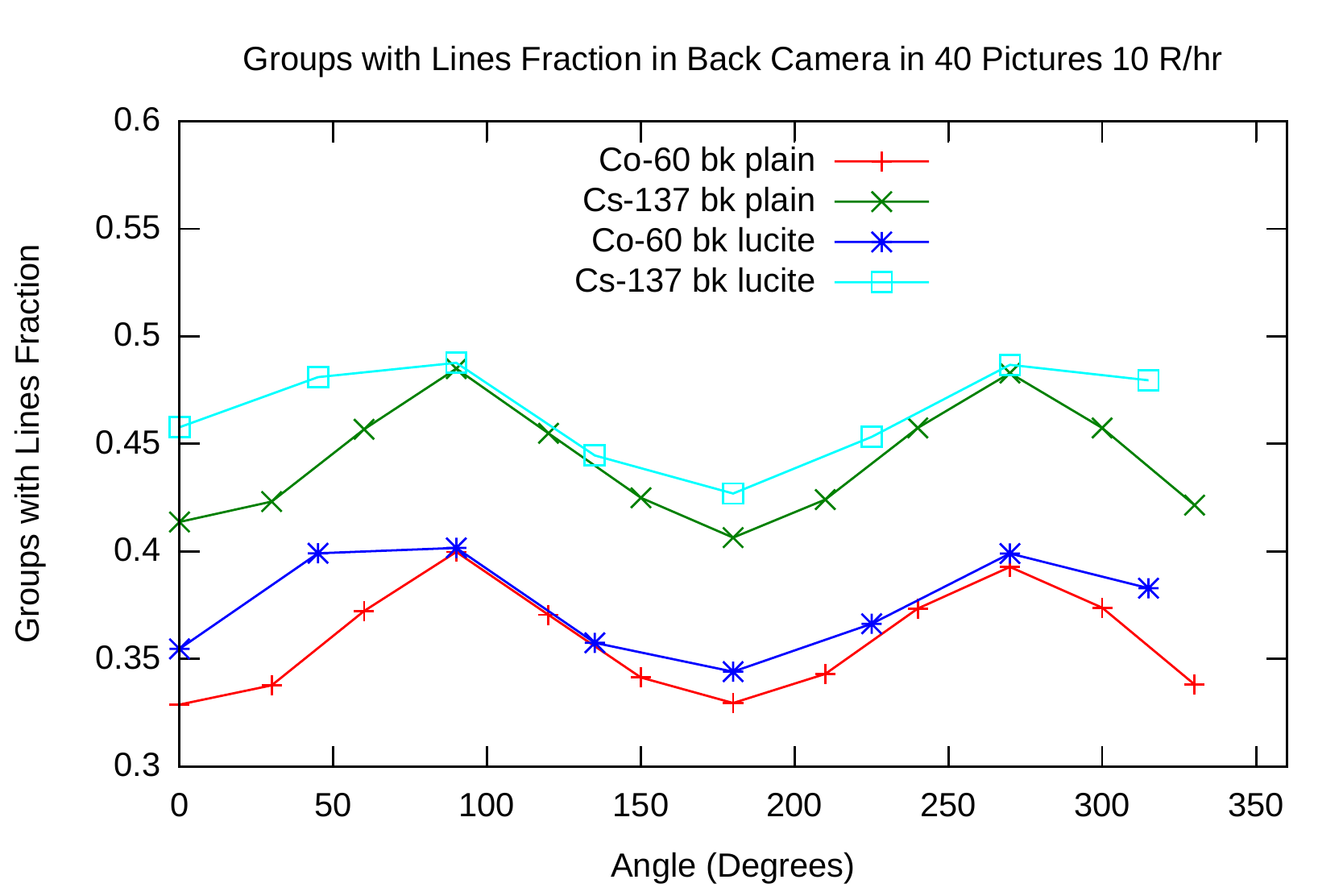}{Nexus S Angular Fraction of Groups with Lines, 10 R/hr, 40 pictures back camera 1.0 sec exposure}{groups_with_lines_fraction_nexus_s_rotation_back_2013_jan_hpil}{0.74}

\myfigures{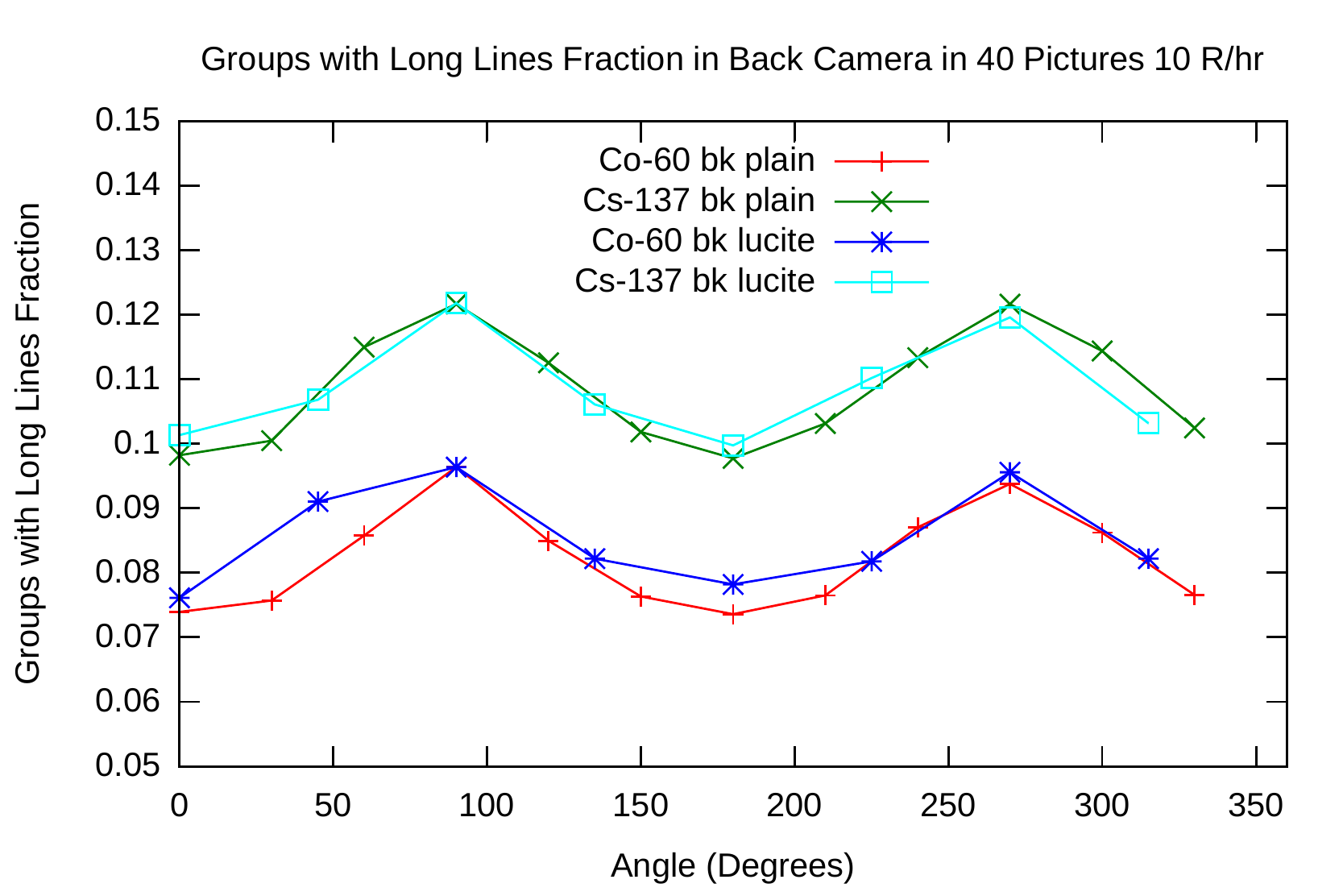}{Nexus S Angular Fraction of Groups with Long Lines, 10 R/hr, 40 pictures back camera 1.0 sec exposure}{groups_with_long_lines_fraction_nexus_s_rotation_back_2013_jan_hpil}{0.74}

\myfigures{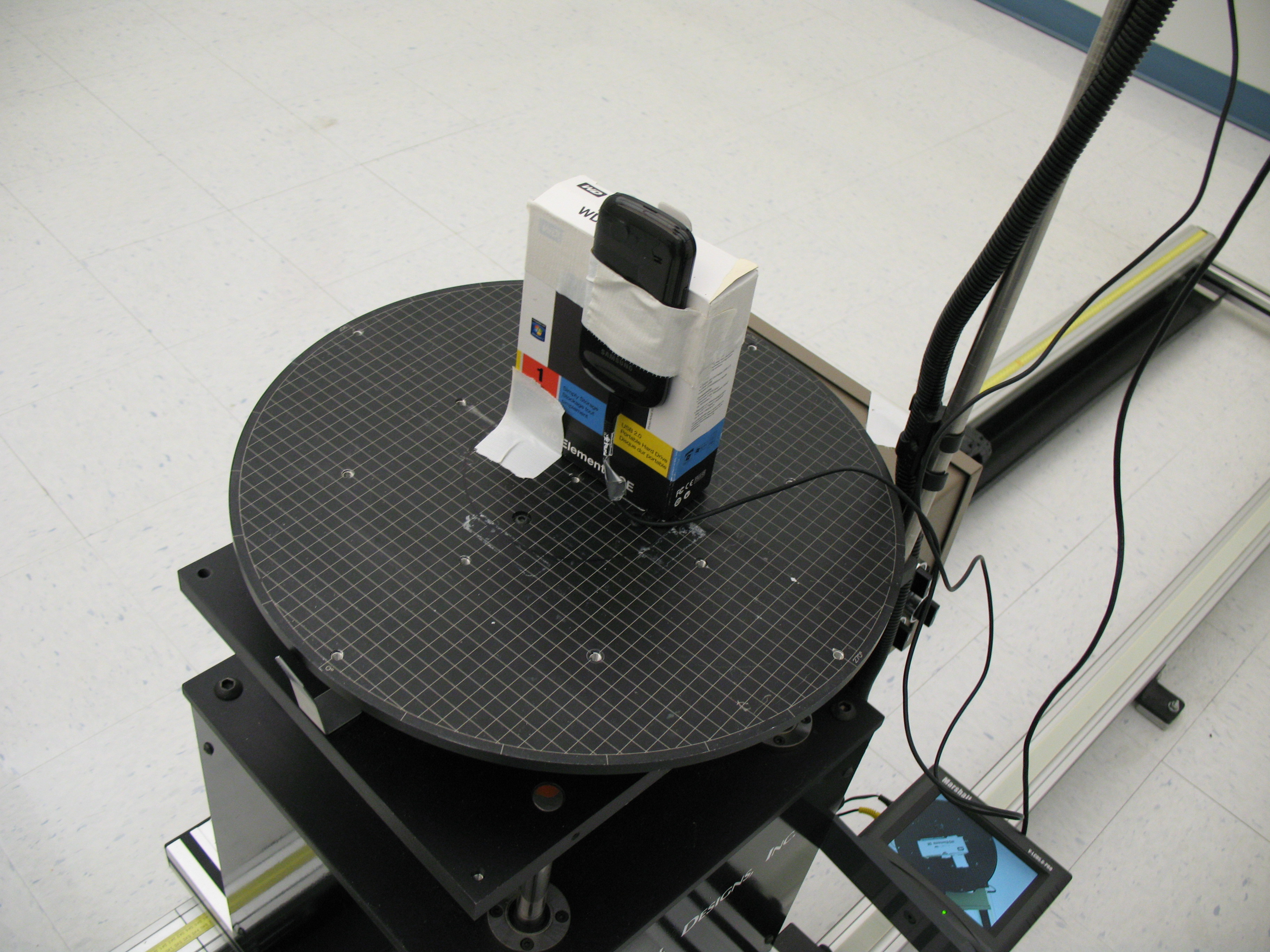}{Nexus S Plain 0 degrees}{nexus_s_plain}{0.08,natwidth=3648,natheight=2736}

\myfigures{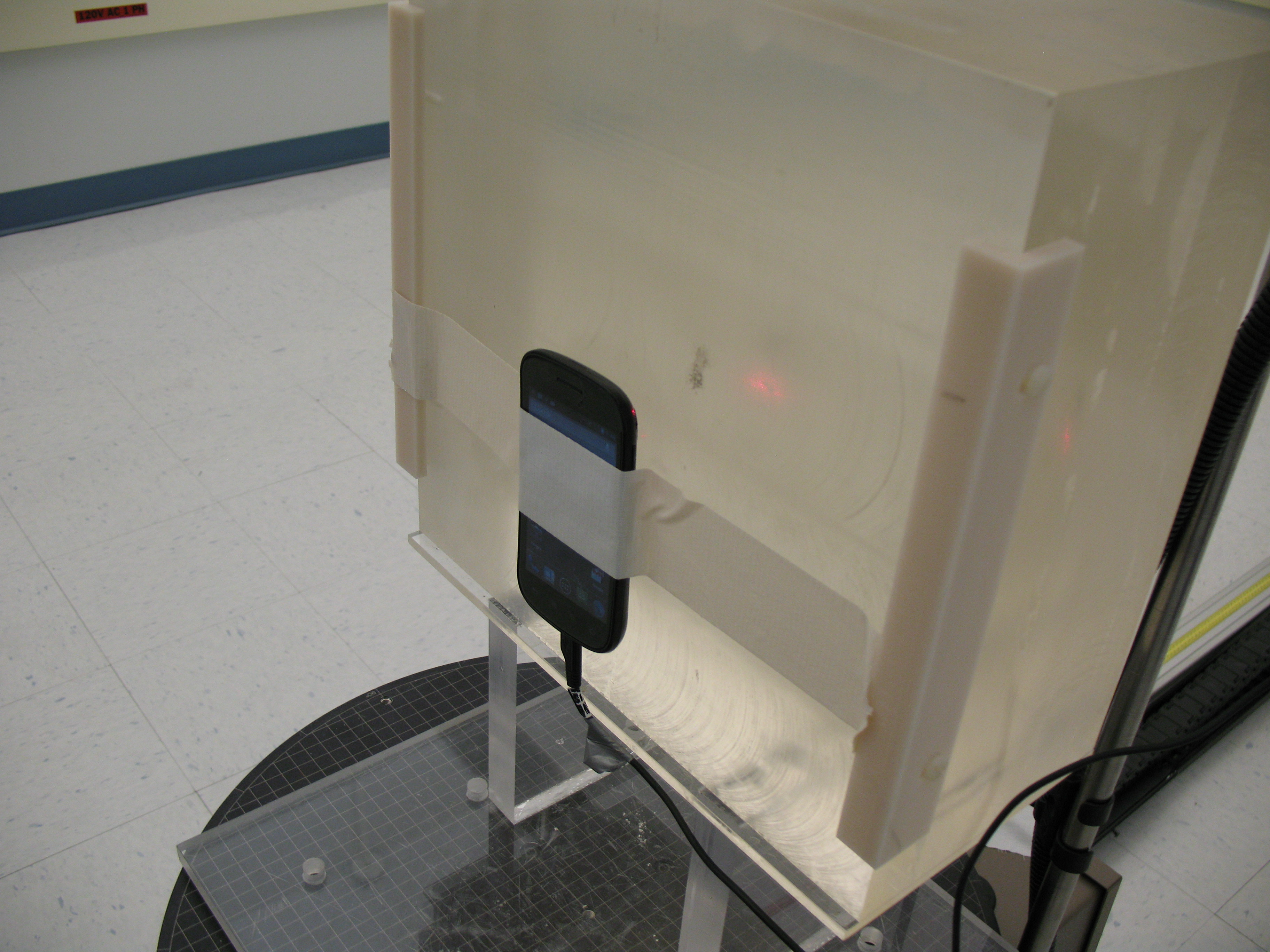}{Nexus S Against Lucite 180 degrees}{nexus_s_against_lucite}{0.08,natwidth=3648,natheight=2736}

For each of the results in this section and the next, 40 pictures were
taken with both the front and the back camera.  Then, a background
picture was generated using the highest value found that was less than
the pixel component's $\bar{x} + 2 \sigma$ as in Equation \ref{m2sig}.
The background picture was then subtracted from each of the 40
pictures, and then line finding and group counting were done on the
images.

Tables \ref{nexus_s_rotation_data} and
\ref{nexus_s_rotation_data_front} provide the summary data for the
four different directions (where 0 degrees is the back camera facing
the source).  The data provided is the total number of groups (See
also Figures
\ref{number_of_groups_nexus_s_rotation_back_2013_jan_hpil} and
\ref{number_of_groups_nexus_s_rotation_front_2013_jan_hpil}), the
number of groups with lines greater than 2 pixels long (GwL) (See also
Figure
\ref{groups_with_lines_fraction_nexus_s_rotation_back_2013_jan_hpil}),
the number of groups with lines greater than 10 pixels long (GwLL)
(See also Figure
\ref{groups_with_long_lines_fraction_nexus_s_rotation_back_2013_jan_hpil}),
and the maximum line length found in all the groups (MLL).  From the
groups data in the tables and in Figures
\ref{number_of_groups_nexus_s_rotation_back_2013_jan_hpil} and
\ref{number_of_groups_nexus_s_rotation_front_2013_jan_hpil}, it can be
seen that the Lucite block attenuates the source strength resulting in
lower numbers of groups compared to plain air near 0 degrees, but
increases the group count from scattering when the cell phone camera
is between or beside the lucite.  The lucite testing was done to
determine the effect of carrying the cellphone next to a human since
the lucite is made of hydrogen, carbon and oxygen.  Depending on
whether the human is between the phone and the source, or the phone is
between the human and the source, the measured dose rate will either
be decreased or increased.  For the in plain air data, the dose
differences when rotated are caused by different attenuation and
scattering from the internal cellphone components.


Examining the fraction of groups with lines (Figure
\ref{groups_with_lines_fraction_nexus_s_rotation_back_2013_jan_hpil})
and the fraction of groups with long lines (Figure
\ref{groups_with_long_lines_fraction_nexus_s_rotation_back_2013_jan_hpil})
leads to important conclusions.  A key detail to note is the fractions
vary with the incoming gamma angle.  At any individual angle, the
fractions of groups with lines are different, so when the incoming
angle is known, the nuclides \coSZ{} and \csOTS{} can be
distinguished. However, because the data nearly overlap between the
\coSZ{} and the \csOTS{} at different incoming angles, if the incoming
gamma angle is not known and cannot be determined, the nuclides cannot
be distinguished. However, the cellphone does have both multiple
cameras and orientation sensors; it is at least conceivable that this
difference in fraction might be able to be used to determine the
direction to the gamma source.  This would require accurate position
registration with the time when the picture was taken, and significant
amounts of data transferred to the server, so it may not be practical.
The ratio between dose detected at the front and back camera might be
usable for direction finding as well, but the lower resolution on the
front camera would require more data to be used.

\section{Dose Results}



Using the same test protocol as the previous section, the response to
different dose rates was tested.  In this case different phones were
placed into gamma fields ranging\footnote{We tried running the Nexus S
  phone at 700,000 mrem/hr, but there were too many crashes in the
  software to be able to complete the test.} from 1 mrem/hr to 100,000
mrem/hr.  As can be seen from Table
\ref{dose_to_number_of_groups_results}, the Nexus S has the highest
groups per mrem/hr per image, which indicates that it is the best for
detecting radiation on a per image basis.  From Figure
\ref{num_groups_per_dose_2013_jan_hpil} there is clear correspondence
between the number of groups and the dose rate.  The figure also shows
that the Galaxy SIII has excessive noise in the camera that is not
compensated for by the existing noise processing in CellRad.  Table
\ref{two_nexus_s_phone_comparison} provides data gathered from two
different Nexus S phones, which indicates that phone to phone
intra-model variance is fairly low, compared to between-model
variance.

\begin{table}[!htb]
  \centering
  \caption{Comparison of two Nexus S phones' group count from data in 10,000 mrem/hr field and one background case}
  \label{two_nexus_s_phone_comparison}
  \begin{tabular}{l|l|l|l}    
    Case	& 3334\ldots & 3231\ldots & \% Change\\
    \hline
    background	&0	&0	& NA\\
    \coSZ{} 0\degree &99868	&102965	&3.0\%\\
    \coSZ{} 90\degree &73267	&75457	&2.9\%\\
    \coSZ{} 180\degree &88264	&90614	&2.6\%\\
    \csOTS{} 0\degree &78143	&78341	&0.3\%\\
    \csOTS{} 90\degree &52068	&55235	&5.7\%\\
    \csOTS{} 180\degree &72336	&73850	&2.1\%\\
  \end{tabular}
\end{table}

\begin{table}[!htb]
  \centering
  \caption{Phones Dose Tested Settings}
  \label{dose_to_number_of_groups_phone_list}
  \begin{tabular}{l|l|l|l}
    Name & Back Camera &  Time (s) & Settings\\
    \hline
    Samsung Nexus S & 2560x1920 & 1.0 &  -m fireworks -e 0\\
    Samsung Galaxy Nexus & 2592x1944 & 0.125 & -m night -e 0\\
    Samsung Galaxy SIII & 3264x2448 & 0.125 & -m night -e 0\\
    LG Nexus 4 & 3264x2448 & 1.0 & -m steadyphoto -e 12\\
  \end{tabular}
\end{table}

\begin{table}[!htb]
  \centering
  \caption{Phones Dose Tested Bright and Bad\_bright Settings}
  \label{dose_to_number_of_groups_phone_list_bright_bb}
  \begin{tabular}{l|l|l}
    Name & Bright & Bad\_bright\\
    \hline
    Samsung Nexus S & 64 & 48\\
    Samsung Galaxy Nexus & 100 & 80\\
    Samsung Galaxy SIII & 48 & 32\\
    LG Nexus 4 & 48 & 32\\
  \end{tabular}
\end{table}

\begin{table}[!htb]
  \centering
  \caption{Groups Per mrem Per image}
  \label{dose_to_number_of_groups_results}
  \begin{tabular}{l|l|l|l|l|l}
    Phone	& average	&min	&max	&1 R/hr \coSZ{}	&1 R/hr \csOTS{}	\\
    \hline
    Nexus S	&0.20295	&0.12708	&0.2753	&0.2753	&0.2106	\\
    Galaxy SIII	&0.13618	&0.01185	&1.1	&0.013	&0.0133	\\
    Nexus 4	&0.01198	&0.009	&0.025	&0.009	&0.012125	\\
    Nexus Galaxy&0.00450	&0	&0.0075	&0.009	&0.004875 \\
  \end{tabular}
\end{table}

\myfigures{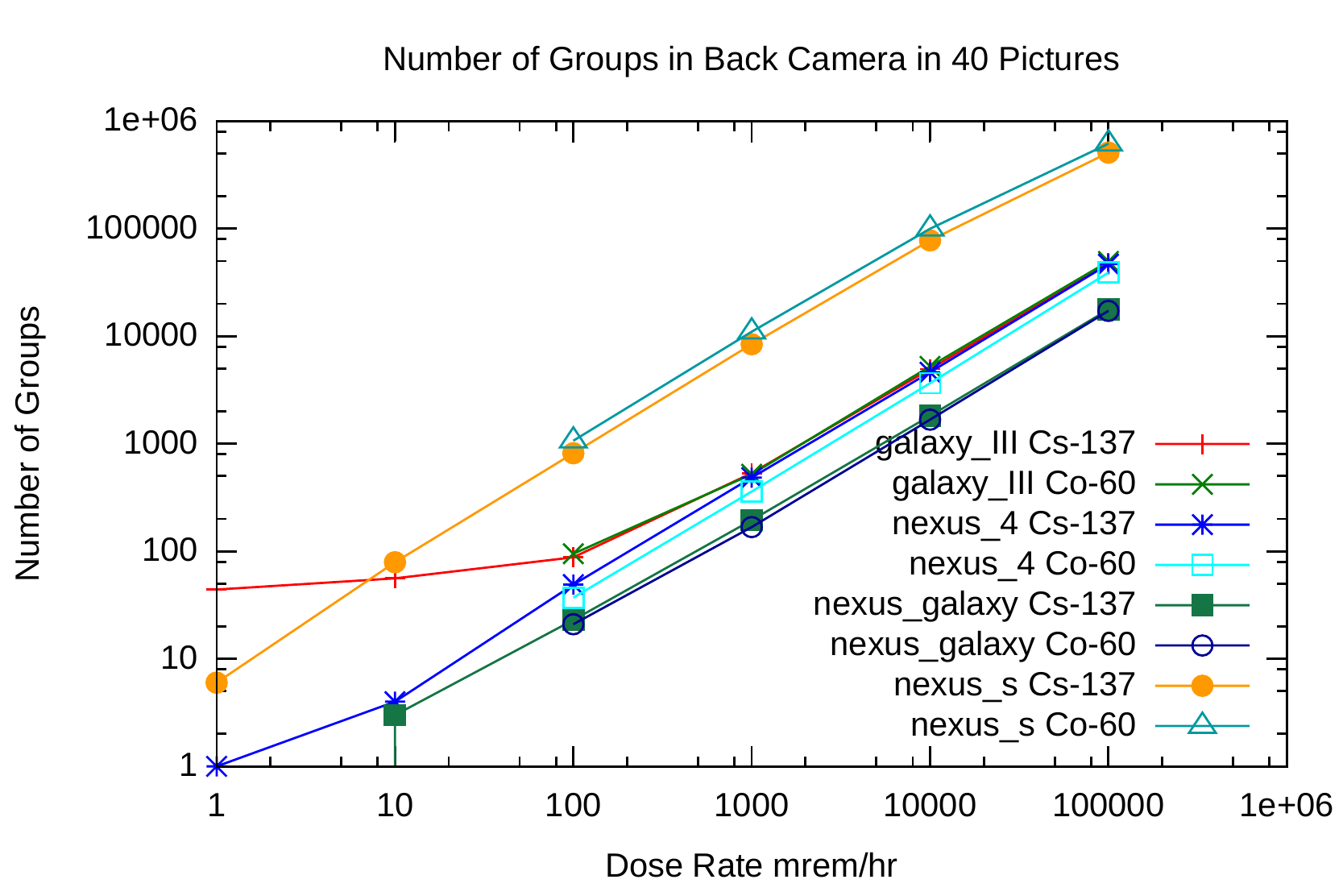}{Dose to Number of Groups}{num_groups_per_dose_2013_jan_hpil}{0.75}

\section{Conclusion}


This project has demonstrated that a smartphone with a camera can be
used as a low sensitivity dose rate meter.  If the radiation is coming
from a known direction, with sufficient data, limited spectrum
information can be determined.  There are better detectors than a
cellphone camera will likely every be, but there are many more
cellphones around than high-quality radiation detectors, which means
there are times when the cellphone can be the best detector available.

\section{Acknowledgments}

This manuscript has been authored by Battelle Energy Alliance, LLC
under Contract No. DE-AC07-05ID14517 with the U.S. Department of
Energy. The United States Government retains and the publisher, by
accepting the article for publication, acknowledges that the United
States Government retains a nonexclusive, paid-up, irrevocable,
worldwide license to publish or reproduce the published form of this
manuscript, or allow others to do so, for United States Government
purposes.

Work supported by the U.S. Department of Energy, Nuclear
Nonproliferation and Security Administration, and the U.S. Department of
Defense, Office of Secretary Of Defense under DOE Idaho Operations
Office Contract DE-AC07-05ID14517.

Many thanks to: Project Leaders Steven H. McCown and Gus Caffrey,
Technical discussions from Shane Hansen, Woo Y. Yoon, Daren R. Norman,
Hope Forsmann, Glenn Knoll, Edward H. Seabury, Thayne C. Butikofer,
Paul K. Halversen, John Richardson, Scott B. Brown and Ryan C. Hruska.
Experimental support and technical discussion from Christopher
P. Oertel, Glenna L. Seal, Norman A. Rhodehouse, Laurel Hill, Byron
Christiansen, Bryce J.  Woodbury, Gary Engelstad, Jennifer A. Turnage,
John R. Giles, and Kent L. Brinker.  In addition to other
contributions, Dr. Carl A. Kutsche suggested that the difference in
fraction of groups with lines with incoming angle could potentially be
used to determine direction to the gamma source.


\myfigures{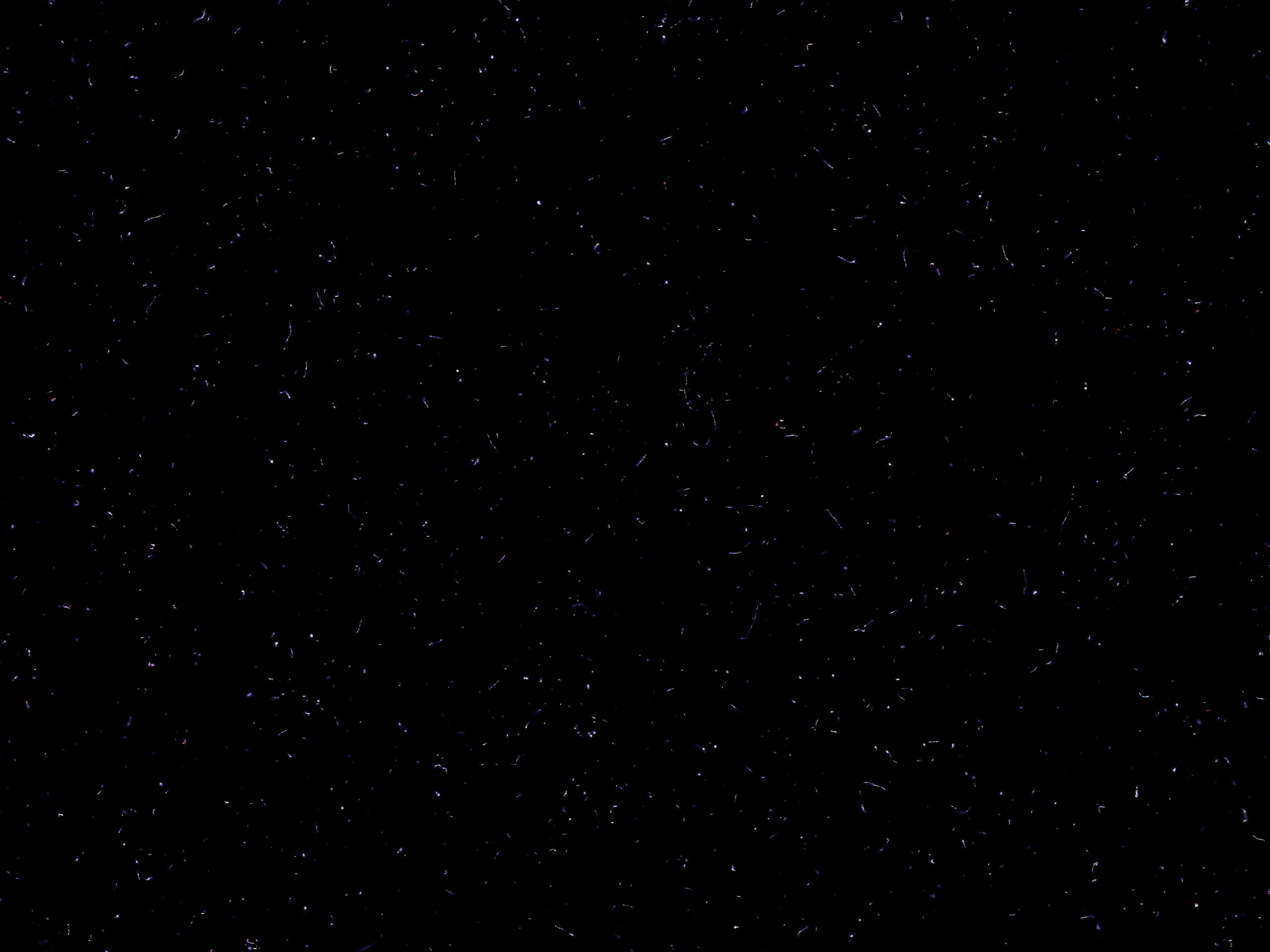}{Example image from \csOTS{}}{2012_aug_16_site_cs-137_long_1_bk_hd_64c}{0.1,natwidth=2560,natheight=1920}

\myfigures{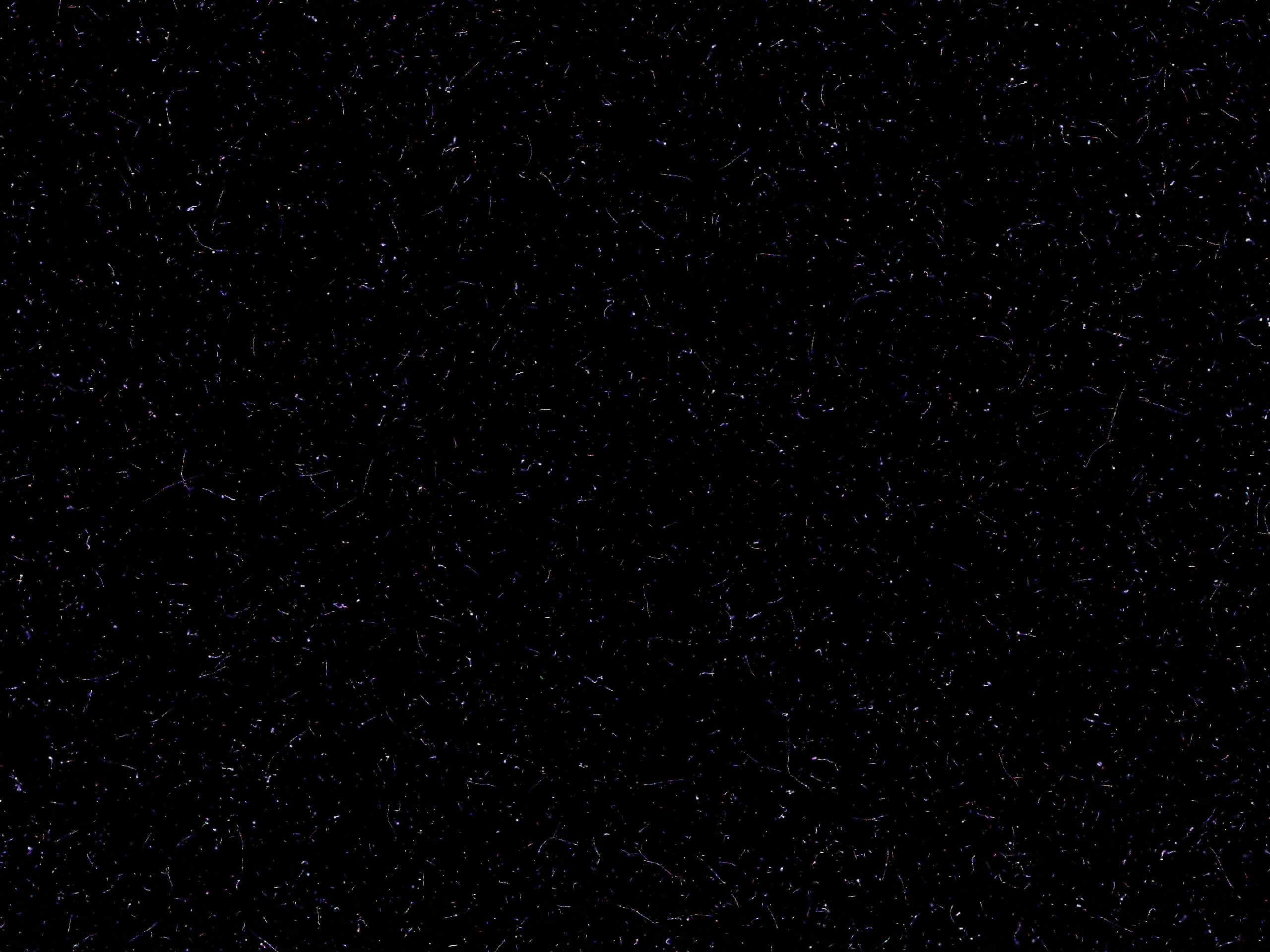}{Example image from \coSZ{}}{2012_aug_16_site_co-60_long_1_bk_hd_64c}{0.1,natwidth=2560,natheight=1920}

\myfigures{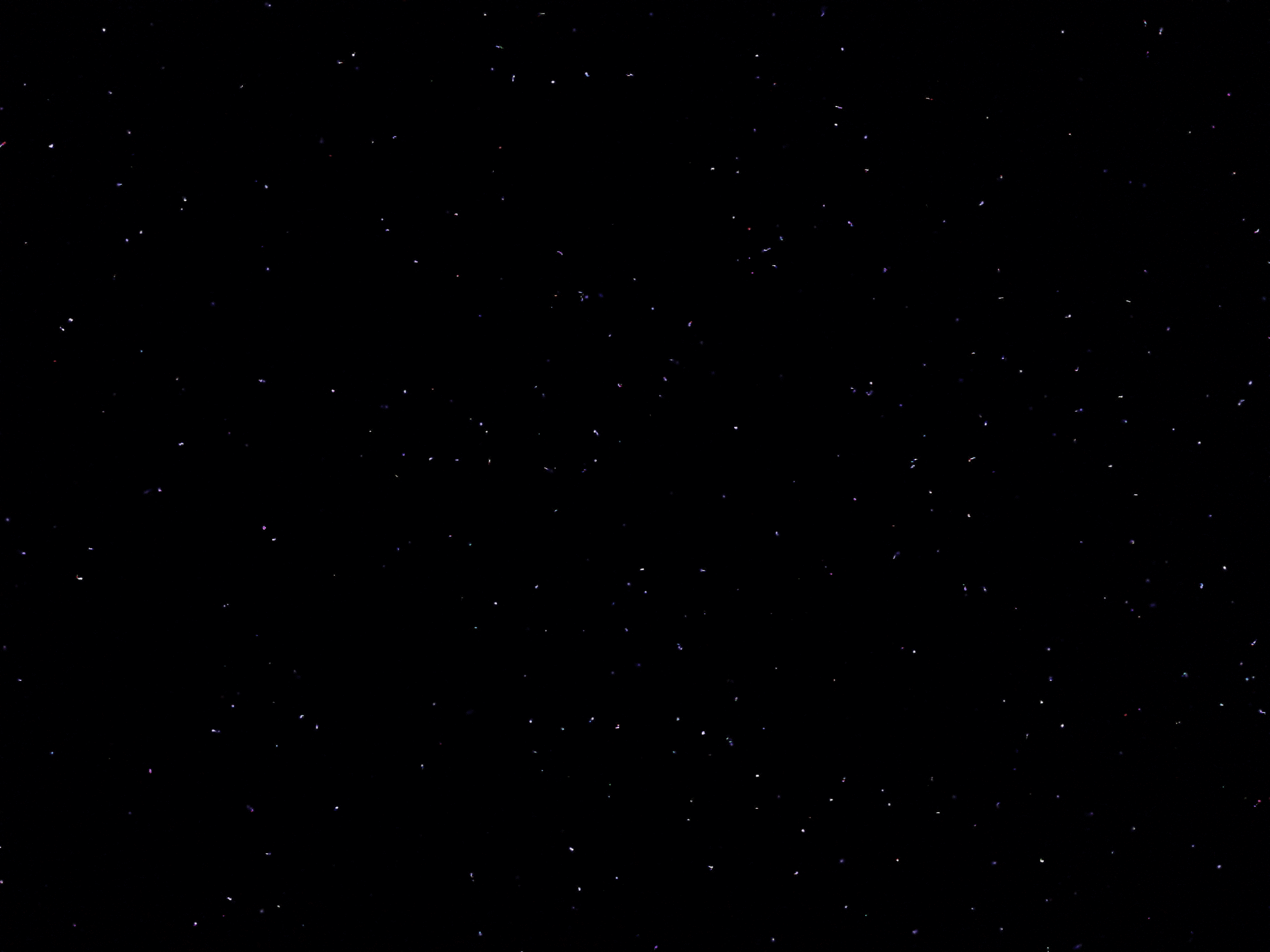}{Example image from \amTFO{}}{2012_aug_16_site_am-241_long_1_hd_64c}{0.1,natwidth=2560,natheight=1920}

\myfigures{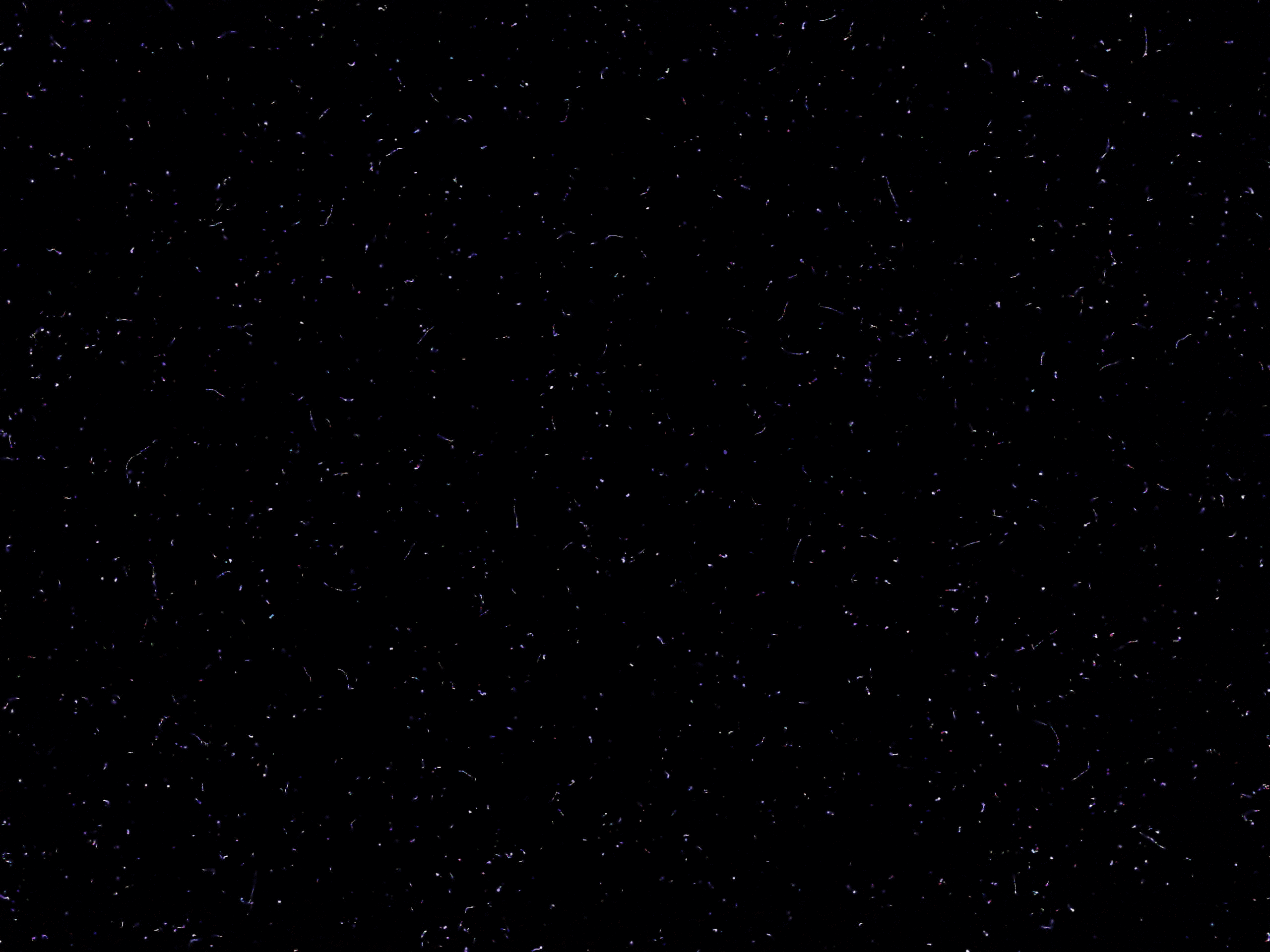}{Example image from \irONT{}}{2012_aug_16_site_ir-192_long_1_hd_64c}{0.1,natwidth=2560,natheight=1920}

\myfigures{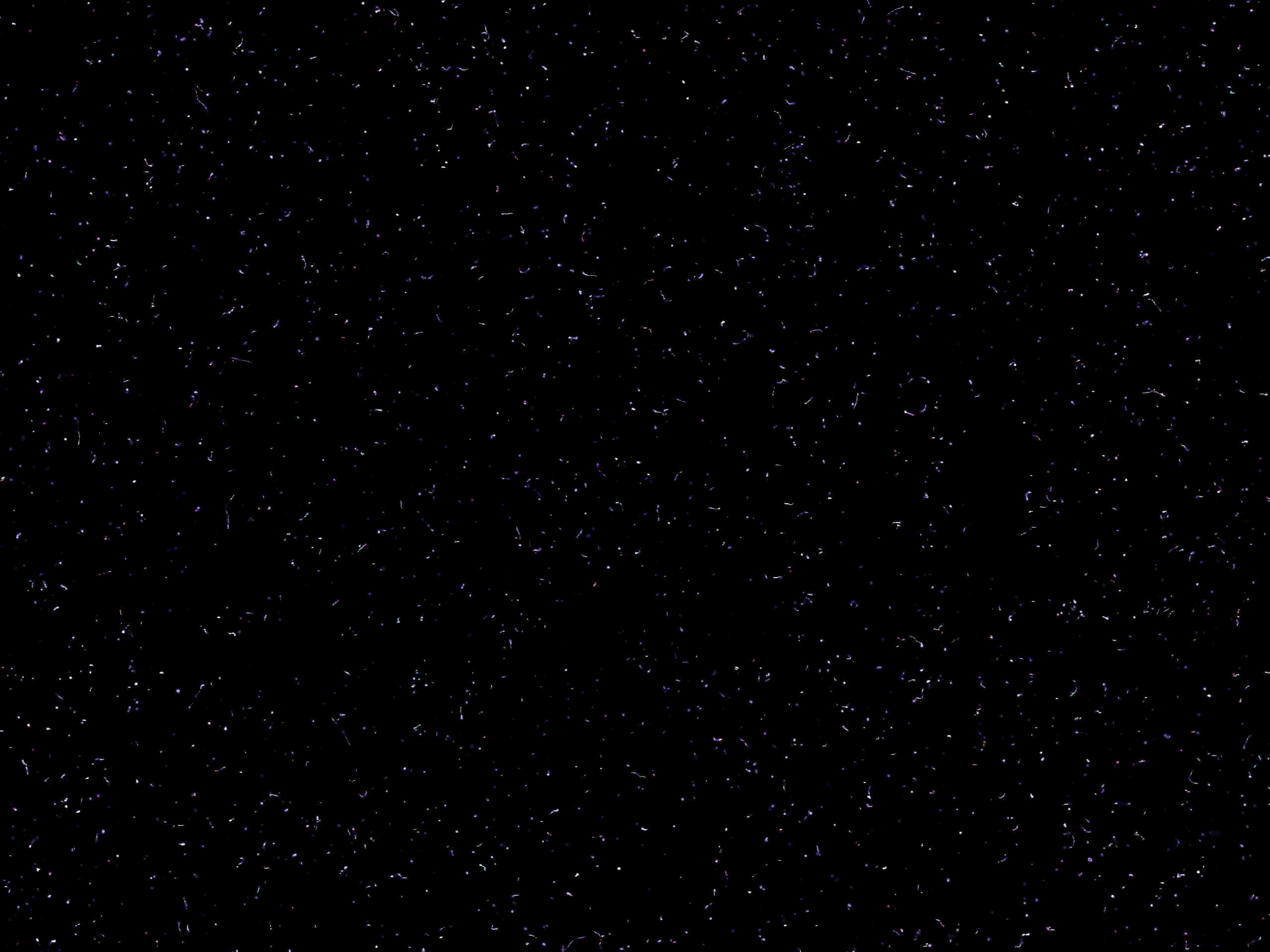}{Example image from \seSF{}}{2012_aug_16_site_se-75_long_1_hd_64c}{0.1,natwidth=2560,natheight=1920}

\end{document}